\documentclass[preprint,12pt]{elsarticle}
\usepackage[utf8]{inputenc}
\usepackage{graphicx, url}
\usepackage{amssymb}
\usepackage{amsmath}
\usepackage{caption}
\usepackage{subcaption}
\usepackage{mathtools}
\usepackage[T1]{fontenc}
\journal{Advanced Powder Technology}

\begin{document}

\begin{frontmatter}

\title{The $\mu(I)$ model and extensions applied to granular material in silo with inserts}

\author[massey]{Samuel K Irvine}
\author{Luke A Fullard}
\author[cant]{Daniel J Holland}
\author[vic]{Daniel A Clarke}
\author[massey]{Thomasin A Lynch}
\author[pyl]{Pierre-Yves Lagrée}

\affiliation[massey]{organization={School of Mathematical and Computational Sciences},
            addressline={Massey University}, 
            city={Palmerston North},
            country={New Zealand}}

\affiliation[cant]{organization={Department of Chemical and Process Engineering},
            addressline={University of Canterbury}, 
            city={Christchurch},
            country={New Zealand}}

\affiliation[vic]{organization={School of Chemical and Physical Sciences, Victoria University of Wellington},
            addressline={PO Box 600}, 
            city={Wellington},
            postcode={6140},
            country={New Zealand}}

\affiliation[pyl]{organization={Sorbonne Université, Institut Jean Le Rond d’Alembert},
            addressline={CNRS, UMR 7190}, 
            city={Paris},
            postcode={75005},
            country={France}}

\begin{abstract}
Granular material is often handled using silos with inserts in industrial processes to prevent ``rat-holing" and similar behaviour, optimise mixing behaviour, and prevent segregation. We study the mass flow rate and the static zones of granular silos using the $\mu(I)$ rheology in the incompressible Navier-Stokes equations. We also extend the $\mu(I)$ model to incorporate pseudo-dilatancy and nonlocal fluidity. We find that insert shape has a significant effect on flow rate and static material. We also find that the effect of inserts on flow rate is not strongly affected by the extensions, while the amount of static material is highly dependent on the interactions between insert shape and extensions. Finally, we find that the effect on flow rate varies with insert size, while the amount of static material is not significantly affected by increasing insert size. These results could provide important insights for optimising silo design.
\end{abstract}

\begin{graphicalabstract}
\includegraphics[width=\textwidth]{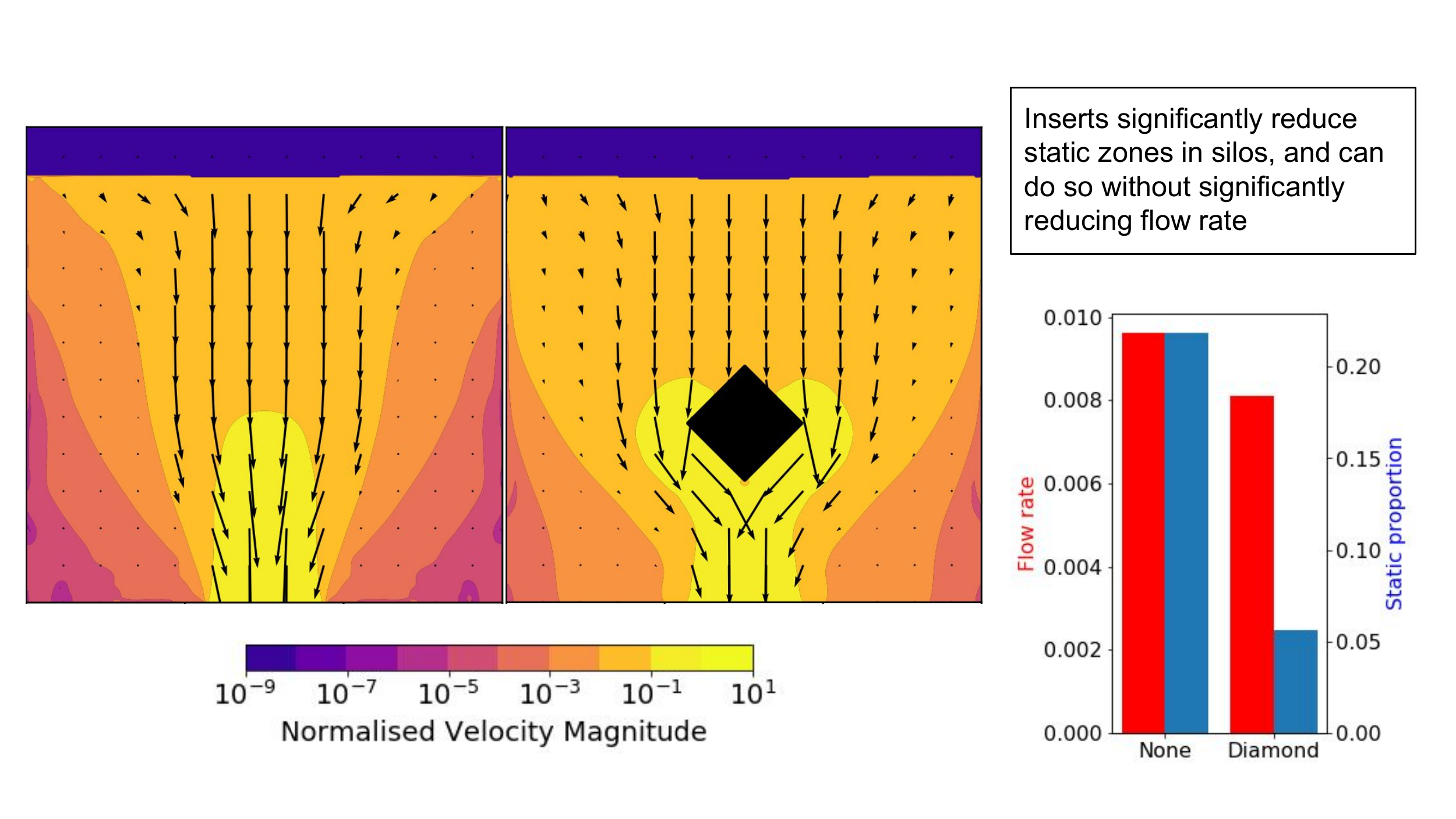}
\end{graphicalabstract}

\begin{highlights}
\item Inserts significantly reduce static and slow moving regions in silos
\item Dilatancy and nonlocal effects maintain similar flow dynamics with inserts
\item Small inserts at the correct height reduce static material without significantly reducing flow rate
\end{highlights}


\end{frontmatter}

\section{Introduction}
Granular material is used for many different industrial processes, and is often stored and processed using silos or silo-like domains. While drawing material from silos several problems can occur, such as jamming, rat holing, and stagnant zones~\cite{gray1997pattern, zuriguel2005jamming}. In order to reduce stagnant zones~\cite{tuzun1985gravity}, change mixing behaviour of the material~\cite{fullard2020mixing}, reduce segregation of mixed material~\cite{cliff2021granular}, or otherwise affect the flow of granular material in a silo, inserts of various shapes are often used. Being able to model the behaviour of granular material in a silo with an insert is highly relevant to many industrial processes.

One method of modelling these flows is modelling each particle individually, called Discrete Element Modelling (DEM)~\cite{cundall1974computer}. By modelling the interaction between each particle, the macroscopic behaviour can be seen in aggregate. This method has previously been used to model the behaviour of a silo with an insert~\cite{hartl2008study, yang2001simulation, kobylka2019loads}, however because DEM requires modelling each particle individually it is computationally expensive. Although many insights can be found using smaller simulations, it is not feasible to model the number of particles used in many real industrial processes.

As an alternative to DEM, granular material can be modelled as a continuous pseudo-fluid. Such a continuum model could capture the desired macro-behaviour of granular flows while reducing the computational overhead involved with modelling the micro-behaviour of granular material. A continuum model capable of accurately replicating the behaviour of granular material in a silo is relevant to many industries, however such a model is difficult to develop due to the many phenomena granular materials exhibit that are difficult to describe.

While multiple continuum approaches have been used~\cite{wojcik2007numerical}, the $\mu(I)$ rheology is the basis for one such continuum model~\cite{gdr2004dense}. The $\mu(I)$ rheology assumes that the frictional behaviour of the material changes depending on the dimensionless inertial number $I$, which determines the flow dynamics in an analogous manner to the Reynolds number for hydrodynamic flows. Using that assumption with the incompressible Navier Stokes equations provides a powerful model which can be applied to many different domains. This work applies the $\mu(I)$ model to a $2D$ silo with an insert, testing whether it is able to capture the expected behaviour when an insert is introduced.

There are some effects that the basic $\mu(I)$ model does not capture. One such effect is compressibility. While the assumption of incompressibility greatly simplifies the numerical calculations, it is known that granular material dilates when sheared. The incompressible $\mu(I)$ model can be extended to capture some dilatancy with a pseudo-compressibility model~\cite{andreotti2013granular} which is described in Section~\ref{sec:dilatancy}. Another factor to consider is that the $\mu(I)$ model is a purely local model, while granular materials show evidence of nonlocal effects: where the flow of material at a point is determined by the behaviour in a neighbourhood around that point rather than just the conditions at that point. However, granular material exhibits nonlocal effects, meaning that the behaviour of the material around a given point can affect the flow at that point. An example of a geometry which displays nonlocal effects clearly is flow down a slope, where there is a difference between the angle at which flow stops when the slope is lowered and the angle at which it starts when the slope is raised~\cite{pouliquen2009non}. This difference exists because the flowing particles agitate their neighbours, maintaining flow when a local model would predict no flow is possible. Another example is an annular shear cell, where local models predict flow sharply going to zero in the areas where the yield criterion is not met, while in experiments an exponential decay is observed~\cite{henann2013predictive}. The $\mu(I)$ model can be extended with a nonlocal fluidity model~\cite{kamrin2015nonlocal}, which spreads out the local flows to capture the nonlocal nature of these flows. This fluidity model is described in Section~\ref{sec:nl}.

\begin{figure}
    \centering
    \includegraphics[width=0.8\textwidth]{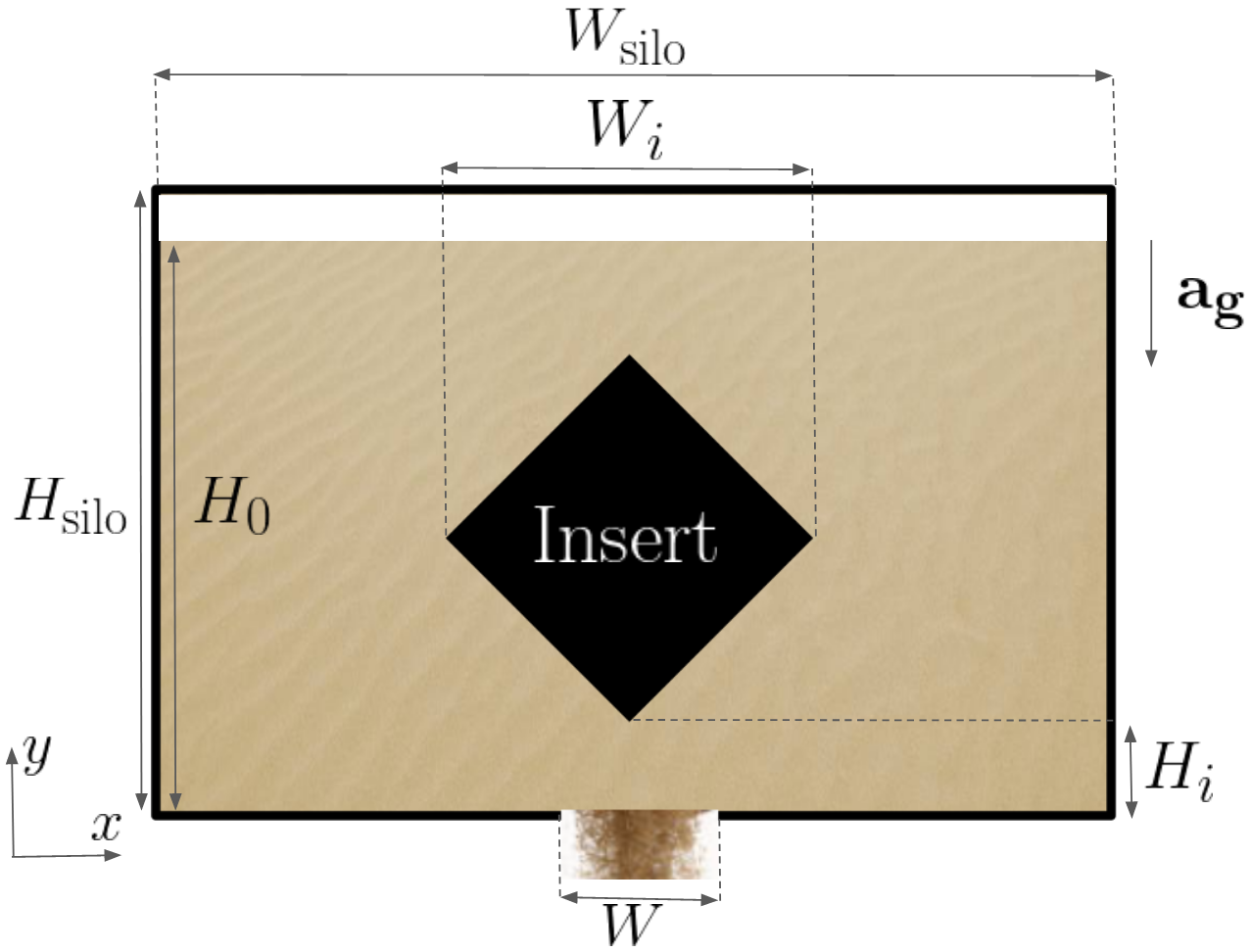}
    \caption{Diagram of the system being studied. Granular material flows around the insert (this case shows the diamond shape) and flows out of the outlet below. The silo is square with $H_\textrm{silo}=W_\textrm{silo}=150d$, and is initially filled to $H_0=135d$. Material is not replaced during the run-time of the simulation.}
    \label{fig:diagram}
\end{figure}

In this paper we model a $2D$ silo with various inserts (as shown in Figure~\ref{fig:diagram}) using the base $\mu(I)$ model and the dilatancy and nonlocal extensions. We examine a silo with no insert and compare to silos with diamond inserts, square inserts, triangular inserts (upwards pointing and downwards pointing), and circular inserts. Examining the diamond insert in more detail we test various insert sizes and  parameter combinations. We test all these different silos for the difference in size and shape of zones where the material is nearly static, as well as the effect on flow rate.

\section{Methods} \label{sec:model}
As per Irvine et~al.~\cite{irvine2022capturing} we describe the $\mu(I)$ rheology, how it applies to the Navier-Stokes equations, and how the dilatancy and nonlocal fluidity extensions are implemented.

\subsection{The $\mu(I)$ model}

The $\mu(I)$ rheology is based on the assumption that the granular friction coefficient varies only on the inertial number $I$, which is given by
\begin{equation}
    I = \frac{|\dot{\gamma}|d}{\sqrt{p/\rho}},
\end{equation}
where $\dot{\gamma_{ij}}$ is the shear rate tensor given by
\begin{equation}
    \dot{\gamma_{ij}} = \frac{\partial u_i}{\partial x_j} + \frac{\partial u_j}{\partial x_i} = 2D_{ij},
\end{equation}
and $|\dot{\gamma}|$ is the second invariant of the shear rate tensor given by
\begin{equation}
    |\dot{\gamma}| = \sqrt{
       \frac{
            \dot{\gamma_{ij}}\dot{\gamma_{ij}}
        }{2}
    } = \sqrt{2D_{ij}D_{ij}},
\end{equation}
$d$ is the particle diameter, $p$ is the pressure, and $\rho$ is the material density.

When the Navier-Stokes equations are combined with the $\mu(I)$ rheology, we obtain the $\mu(I)$ model which was described by~\cite{gdr2004dense} and extended to $3D$ by~\cite{jop2006constitutive}. We assume incompressibility for the base model, using the incompressible Navier-Stokes Equations~(\ref{eq:Navier_Stokes}) which take the form
\begin{equation}\label{eq:Navier_Stokes}
\begin{split} 
&\partial_t\mathbf{u}+\mathbf{u}\cdot\nabla\mathbf{u} = 
\frac{1}{\rho}\left[-\nabla p + \nabla\cdot(2\eta\mathbf{D})\right] -
\mathbf{a_g},\\
&\nabla\cdot\mathbf{u} = 0,
\end{split}
\end{equation}
where $\mathbf{u}$ is the velocity vector, $\rho=\phi\rho_\textrm{granular}+(1-\phi)\rho_\textrm{air}$ is the density derived from the packing fraction $\phi$ and the mix of material and gas density ($\rho_\textrm{granular}$ and $\rho_\textrm{air}$ respectively), $\mathbf{D}$ is the strain rate tensor given by $\mathbf{D}=[\nabla\mathbf{u} + (\nabla\mathbf{u})^T]/2$, and $\mathbf{a_g}$ is the acceleration due to gravity.

The relationship between $\mu$ and $I$ is not fixed, so many different formulations are possible~\cite{kamrin2017hierarchy}. In this paper we make the assumption that the rheology is a kind of Coulomb friction with a coefficient of the form
\begin{equation} \label{eq:muI}
    \mu(I) = \mu_s + \frac{\Delta\mu}{(I/I_0+1)},
\end{equation}
where $\mu_s$, $\Delta\mu$, and $I_0$ are fitting parameters given in Table~\ref{tab:parameters}. Equation~\ref{eq:muI} defines the friction, which is implemented in the Navier-Stokes solver as an effective viscosity defined by
\begin{equation}\label{eq:effective_viscosity}
    \eta = \frac{\mu(I)p}{|\dot{\gamma}|}.
\end{equation}

In order to solve the incompressible Navier-Stokes equations with the varying $\eta$ we use the numerical scheme first implemented with Gerris~\cite{popinet2003gerris} in~\cite{lagree2011granular}, and later with the framework Basilisk~\cite{basilisk, zhou2017experiments, zou2020discharge, zou2022nonsteady}. The Navier-Stokes Equations can be transformed using a projection method into a Poisson equation and a Helmholtz equation. The simulation uses a volume of fluid method representing the granular material as well as the air~\cite{lopez2015electrokinetic}. The boundary conditions are no-slip at the silo walls, silo base, and insert boundary, and zero pressure at the top of the silo and inside the opening.

Another complication posed when implementing the $\mu(I)$ model is the possibility of diverging $\eta$. According to Equation~\ref{eq:effective_viscosity} a static region of material corresponds to an unbounded viscosity, and in a practical flat-bottomed silo there are some areas where there will be zero flow. To avoid divergent viscosity, a maximum viscosity $\eta_\textrm{max}$ is enforced, i.e. the finite $\eta^*$ is used, given by $\eta^* = min(\eta,\eta_\textrm{max})$, where $\eta$ is the viscosity calculated from Equation~\ref{eq:effective_viscosity} and $\eta_\textrm{max}$ is a large constant. This leads to a small creeping flow in these regions which should be static in a physical silo. For our simulations we use $\eta_\textrm{max} = 1800 \sqrt{\rho^2d^5a_g^{-1}}$, which is sufficiently large such that the creeping flow is negligible compared to the flow from the silo draining due to gravity.

It should be noted that this model assumes that the stress and strain rate tensors are aligned. The $\mu(I)$ model relies on visco-plastic theory which assumes that these tensors are in the same direction, which is not always the case~\cite{cortet2009relevance}. Also, the model is not well-posed, so may fail for some parameters and some domains~\cite{barker2015well, heyman2017compressibility}. The parameters and domain studied in this paper do not display evidence of numerical instabilities, however caution should be applied when using these models.

In order to avoid applying different boundary conditions within the length of a single cell, all lengths are chosen to be some integer multiple of the cell width. Due to the tree-based discretization implementation used~\cite{van2018towards}, the cell width is the domain width divided by $2^n$, where $n$ determines the resolution of the simulation and is chosen to be $n=8$ for this work.

\begin{table}[t]
\centering
\begin{tabular}{|l|l|l|l|l|}
\hline
Parameter & Value \\ \hline
Relative density of air $\rho_\textrm{air}/\rho_\textrm{granular}$ & $1.7\times10^{-3}$ \\ \hline
Orifice width $W$ & $9.375d$  \\ \hline
Domain width $W_\textrm{silo}$ & $150d$ \\ \hline
Domain height $H_\textrm{silo}$ & $150d$ \\ \hline
Initial fill height $H_\textrm{silo}$ & $135d$ \\ \hline
Static friction $\mu_s$ & $0.62$ \\ \hline
Friction differential $\Delta\mu$ & $0.48$\\ \hline
Inertial number scaling $I_0$ & $0.6$ \\ \hline
Maximum solid fraction $\phi_\textrm{max}$ & $0.6$\\ \hline
Minimum solid fraction $\phi_\textrm{min}$ & $0.2^*$ \\ \hline
Solid fraction gradient $\phi_\textrm{grad}$ & $0.2^*$\\ \hline
Nonlocal model strength $A$ & $0.5^*$\\ \hline
\end{tabular}
\caption{Parameters used throughout this paper, with lengths being given as multiples of the particle diameter $d$. An asterisk indicates the parameter is only relevant when the appropriate extension is enabled.}
\label{tab:parameters}
\end{table}

\subsection{Dilatancy} \label{sec:dilatancy}

While the assumption of incompressibility greatly simplifies the numerical methods, in reality granular material is known to dilate when sheared~\cite{andreotti2013granular, reynolds1885lvii}. Dilatancy introduces compressibility. Models that fully take into account the compressible Navier-Stokes equations are complex~\cite{barker2017well, pailha2009two, bouchut2021dilatancy}. However for gravity confined flows over the range of inertial numbers relevant to a silo, the packing fraction seems to linearly decrease with the inertial number~$I$~\cite{hurley2015friction}. We can implement this linear dependence as a simple model for dilatancy, with 
\begin{equation} \label{eq:linear_dilatancy}
    \phi = \textrm{max}(\phi_\textrm{max}-\phi_\textrm{grad}I, \phi_\textrm{min})
\end{equation}
where $\phi_\textrm{grad}$ is the linear gradient parameter, $\phi_\textrm{max}$ is a constant representing the packing fraction for a material that is not being sheared ($0.6$ is used throughout this paper), and the minimum packing fraction is set to $\phi_\textrm{min}=0.2$, in order to prevent non-physical negative packing fractions. While other formations of the relationship of $\phi$ to $I$ are sometimes used~\cite{robinson2021evidence}, for simplicity we limit our scope to only consider linear dependence. This packing fraction is used to calculate the bulk density of the material. With non-homogeneous density the incompressible assumption $\nabla\cdot u = 0$ is replaced with a `source term', which is derived from conservation of mass. This source term is given by
\begin{equation}
    \nabla \cdot \mathbf{u} = \frac{1}{\rho}\left ( \frac{\partial\rho}{\partial t} + \mathbf{u}\cdot\nabla\rho \right ),
\end{equation}
where $\rho$ is the bulk density.

The bulk density is dependent both on the material density as well as the packing fraction of the granular material, determined by the linear dilatancy model in Equation~\ref{eq:linear_dilatancy} (if $\phi$ is constant the incompressible assumption is recovered). The source term dilatancy model approximates the source term and incorporates it in the projection method.

\subsection{Nonlocal effects} \label{sec:nl}
The base $\mu(I)$ model and the model with dilatancy are both local models, meaning that the properties of flow at a point are determined only by other properties at that point. In order to account for any nonlocal effects we use a granular fluidity model~\cite{kamrin2012nonlocal, salvador2017modeling, faroux2021coupling}. This model finds the local `fluidity' (which can be conceptualised as an inverse viscosity), and then `spreads out' the fluidity into nearby regions, representing the agitation caused by flowing particles in a neighbourhood that create nonlocal effects. The fluidity $g$ (not to be confused with gravity $a_g$) is related to the $\mu$ value by the relation $|\dot{\gamma}| = \mu g$. The fluidity has some local value $g_l$, which corresponds to the fluidity if there were no nonlocal effects. The local $g_l$ is then `spread out' by the Laplacian term
\begin{equation}
    g = g_l + \xi(\mu(I))^2 \nabla^2 g,
\end{equation}
where $\xi(\mu(I))$ is given by
\begin{equation}
    \xi(\mu(I)) = A \sqrt{\frac{\mu_s + \Delta\mu - \mu(I)}{\Delta\mu(\mu(I)-\mu_s)}} \label{eq:non_local}
\end{equation}
where $A$ is a parameter which determines how strong the nonlocal effects are, with $A = 0$ corresponding to a purely local model. With the fluidity $g$ calculated, the relations $|\dot{\gamma}| = \mu g$ and $\eta = \frac{\mu P}{\dot{\gamma}}$ can be combined to give the effective viscosity as
\begin{equation}
    \eta = \frac{P}{g},
\end{equation}
which is used in Equation~\ref{eq:Navier_Stokes} in the same manner as the base $\mu(I)$ model. Boundary conditions are zero fluidity $g=0$ at the silo base and walls (corresponding to infinite viscosity i.e. no-slip) and zero normal flux $g_n = 0$ at the opening and top. For the insert boundary conditions, practically we found that using a small fluidity (i.e. $g=10^{-3}\sqrt{a_gd}$) boundary condition was sometimes necessary for stability. This could potentially create a kind of fluidisation similar to fluidised beds~\cite{goldschmidt2001hydrodynamic}. However, due to the small value of fluidity used and the low sensitivity we found to this value, fluidisation does not seem to be occurring. Boundary conditions for granular flow, particularly nonlocal fluidity, is an area which required additional research and we omit further study of them from this paper.

\section{Results}

\subsection{Insert shape}
\begin{figure}
    \centering
    \begin{tabular}{ccc}
        \includegraphics[scale=0.45]{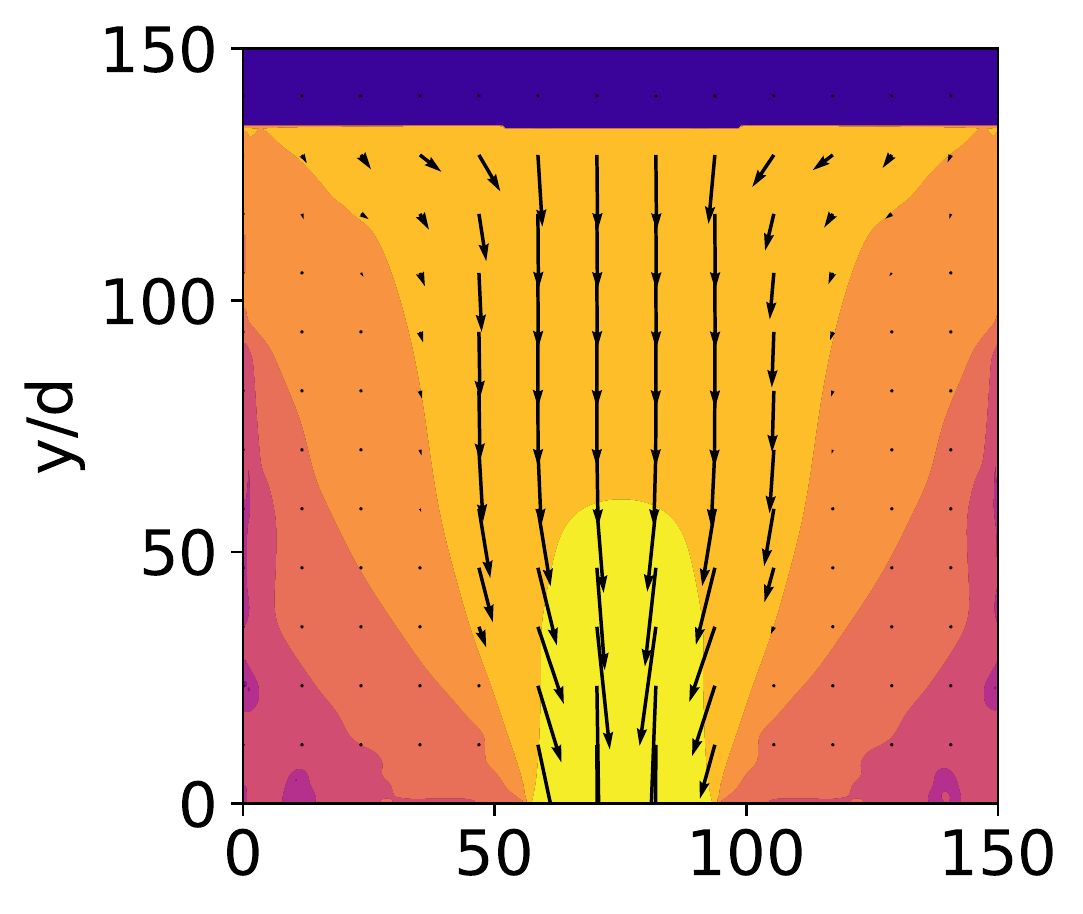} &
        \includegraphics[scale=0.45]{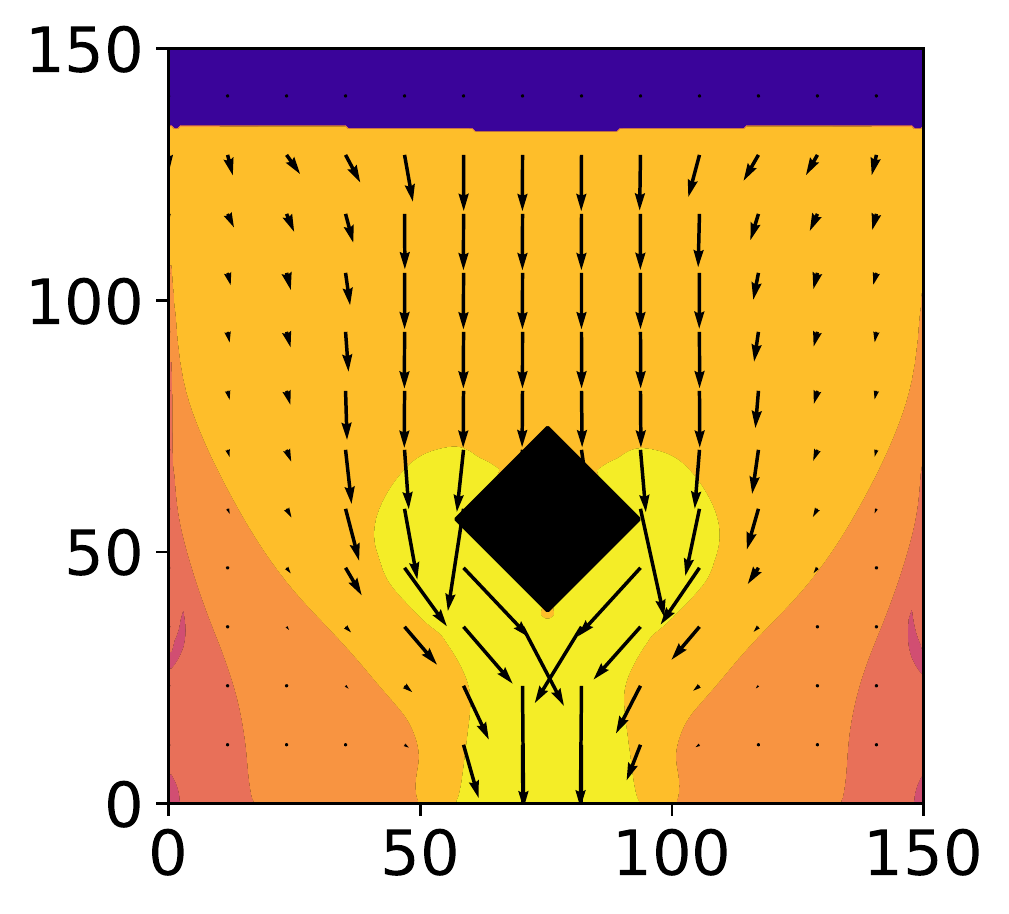} &
        \includegraphics[scale=0.45]{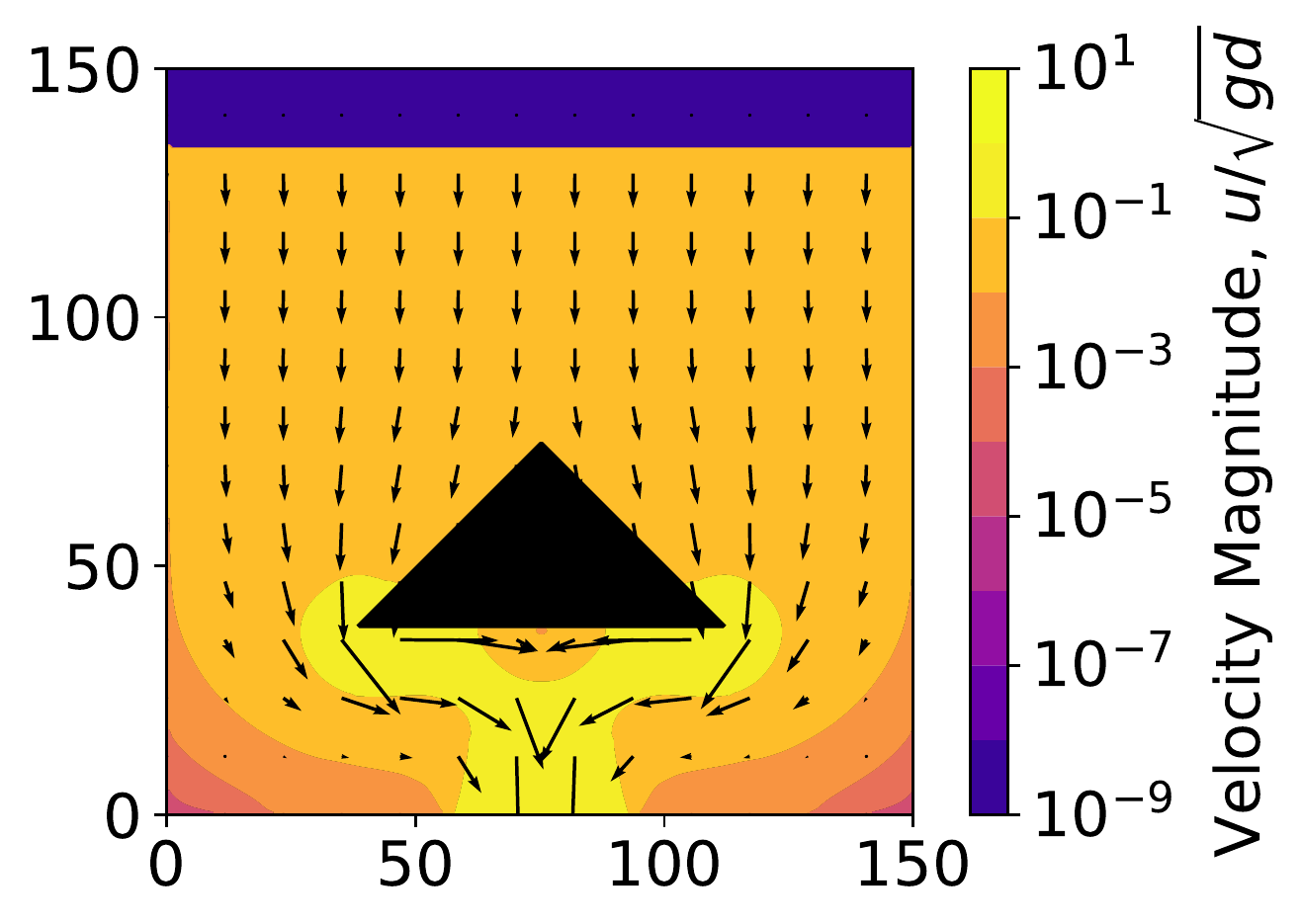} \\
        No insert & Diamond insert & Upwards triangle insert\\[6pt]
        \includegraphics[scale=0.45]{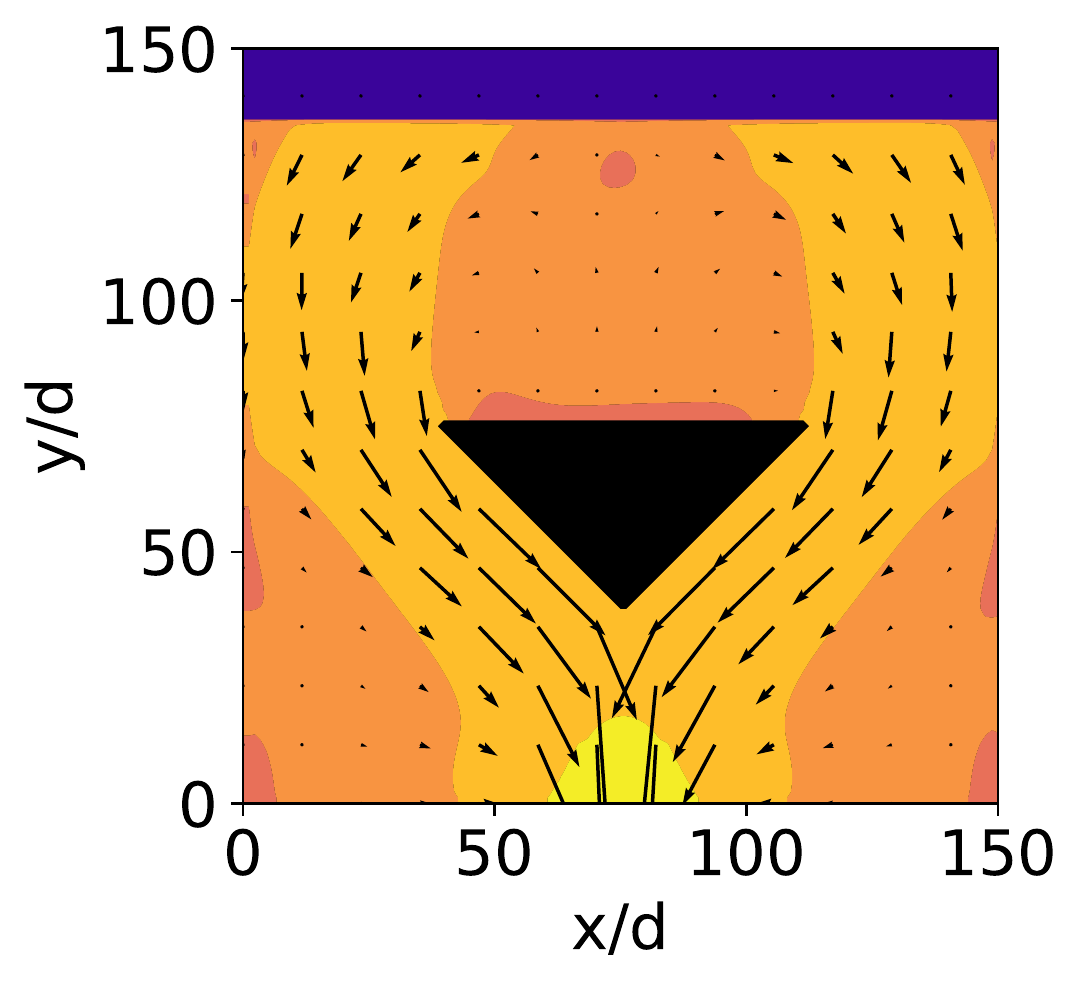} &
        \includegraphics[scale=0.45]{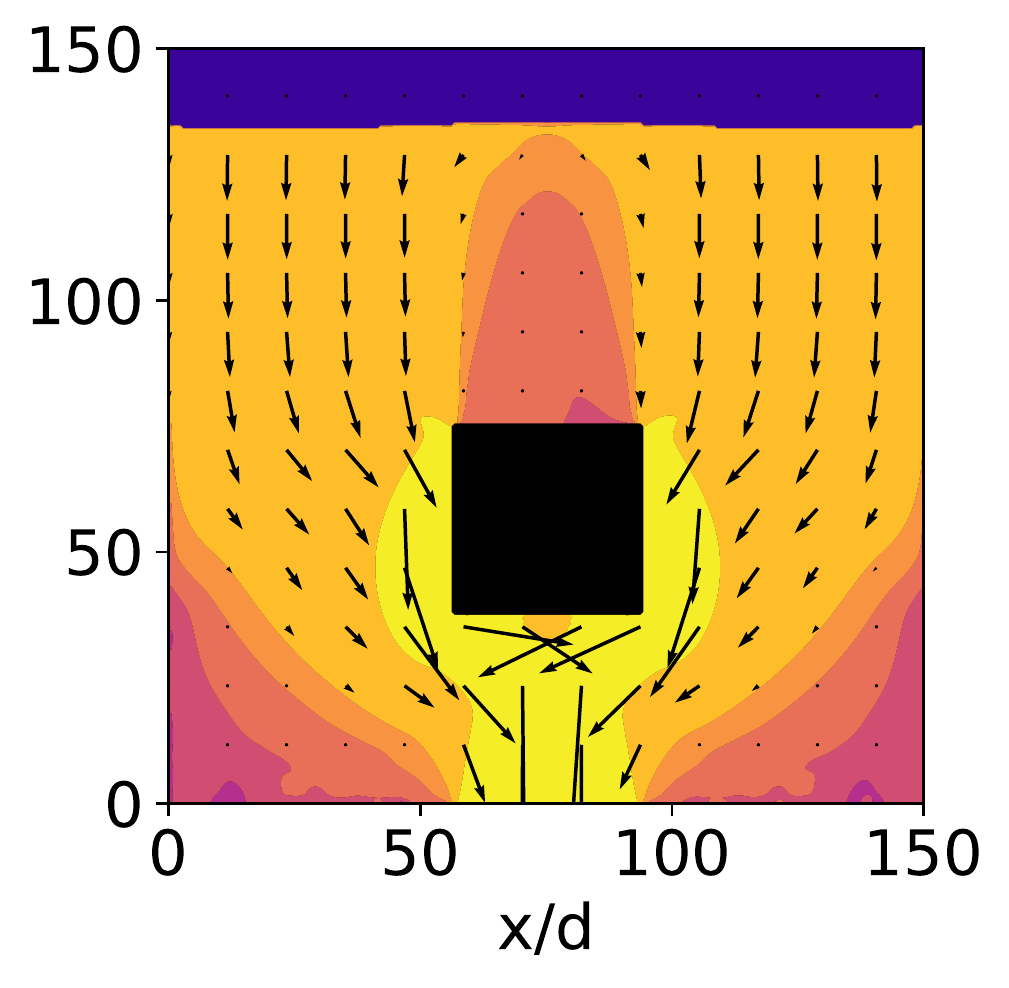} &
        \includegraphics[scale=0.45]{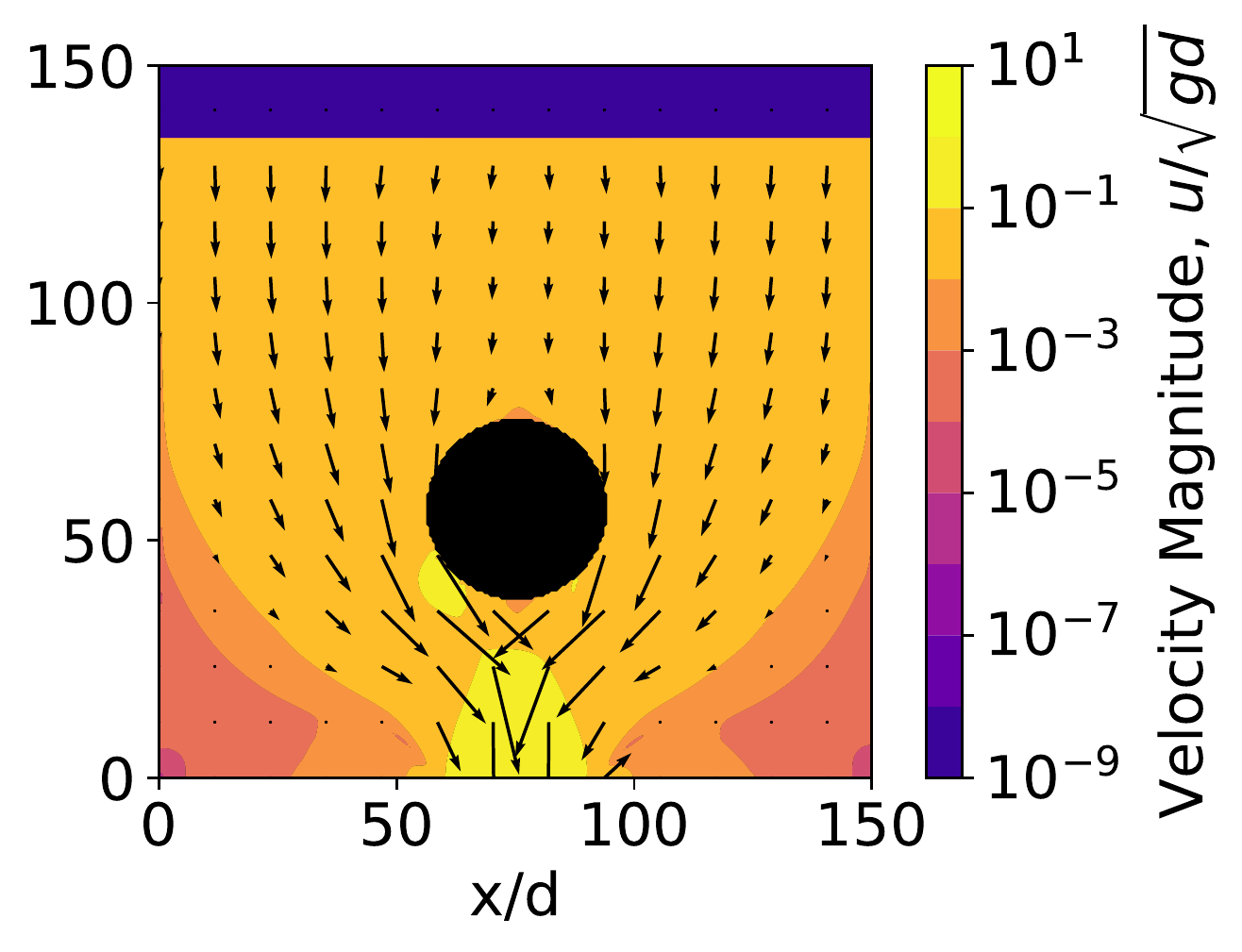} \\
        Downwards triangle insert & Square insert & Circle insert\\[6pt]
    \end{tabular}
    \caption{Velocity in silos for various different insert shapes. The square, diamond, and circle inserts are sized to have height and width equal to the width of the silo opening, with the triangle inserts having the same height and twice the width.}
    \label{fig:shape_contour}
\end{figure}

The geometry of the insert has a large influence on the flow behaviour, so we first apply the base $\mu(I)$ model to a variety of differently shaped inserts, looking at square, diamond, triangular (upwards pointing and downwards pointing), and circular inserts, as well as comparing the silo with no insert. Each insert is sized such that the height remains consistent, rather than the area (which compared to the square insert is reduced by a factor of $4/\pi$ for circle inserts and by $2$ for a diamond insert) or width of insert (which is doubled for both triangular inserts when compared to the square insert). Figure~\ref{fig:shape_contour} shows velocity contours around various different shapes of insert. The case with no insert exhibits `funnel flow' behaviour, with a fast moving region in the center and large slow moving zones to the sides. The circle, diamond, and triangle inserts all disrupt the funnel area which spreads the flow out into otherwise near-static regions, with the circle appearing to have the largest effect and the diamond the smallest.

It should be noted that the flat bottoms of the square and the triangle inserts may not display realistic results directly underneath the insert. In a physical silo with a flat-bottomed insert, we would expect a void zone with little to no material. The unrealistic rapid horizontal flow directly beneath the insert is likely an artifact of the continuum model not being designed to model these void zones.

\begin{figure}
    \centering
    \begin{tabular}{ccc}
        \includegraphics[scale=0.45]{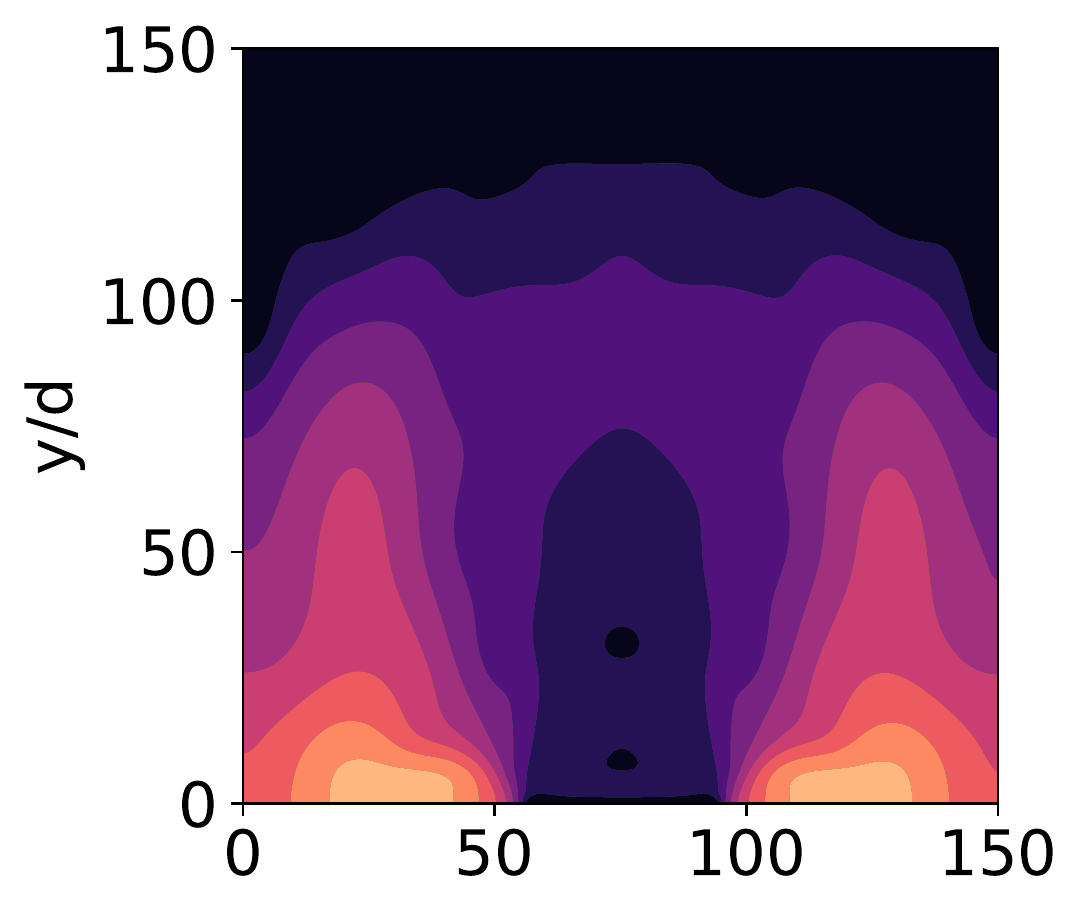} &
        \includegraphics[scale=0.45]{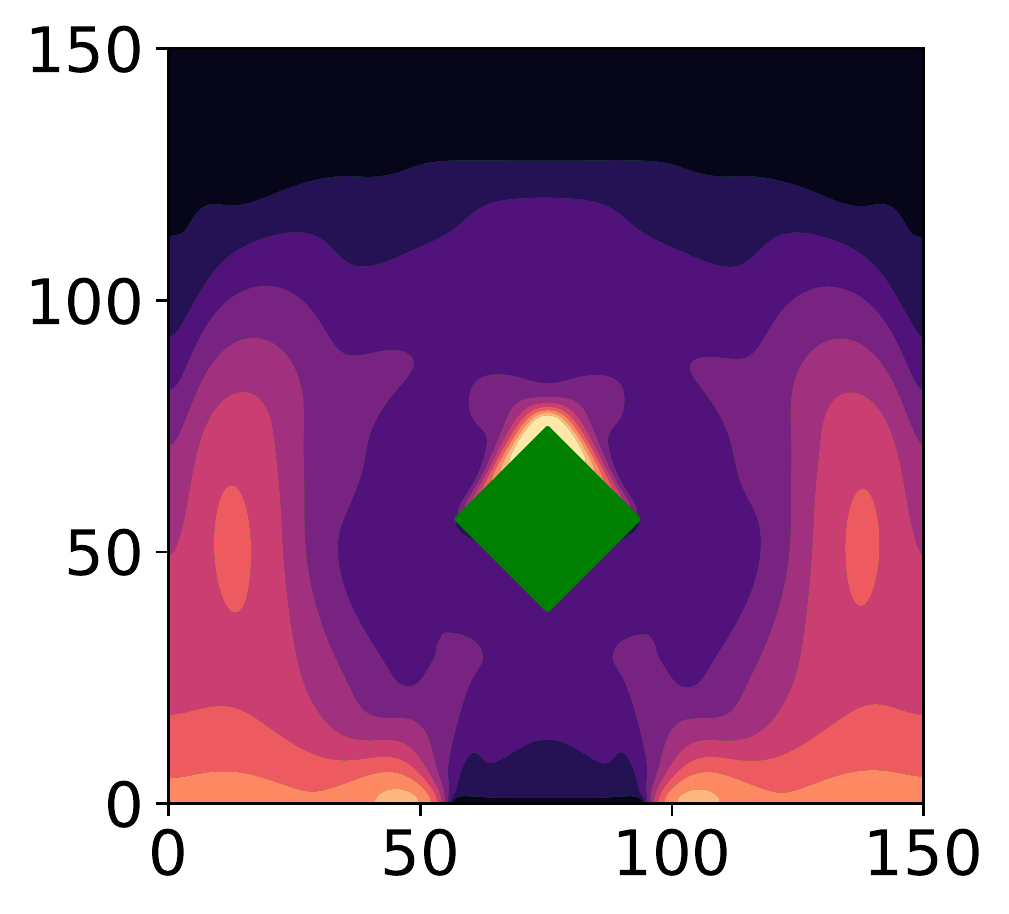} &
        \includegraphics[scale=0.45]{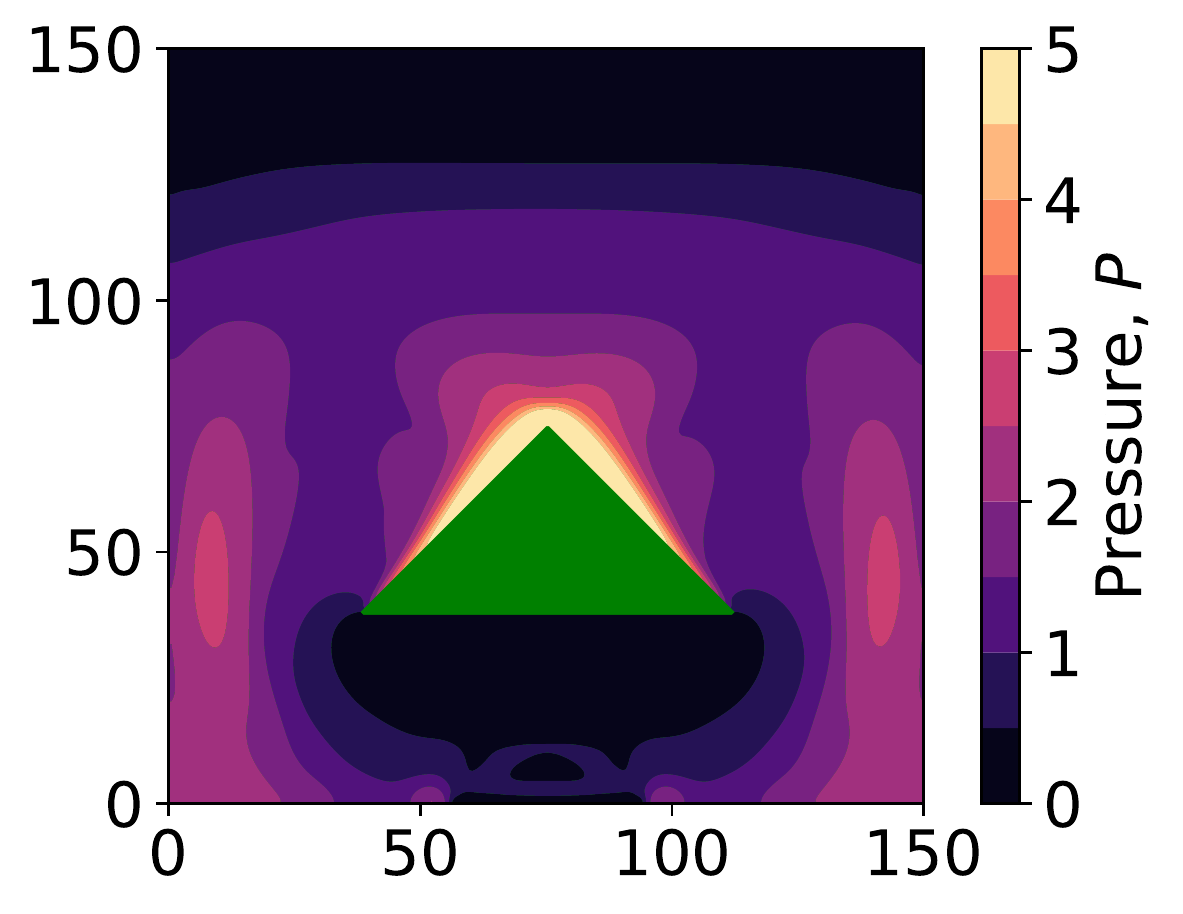}\\
        No insert & Diamond insert & Upwards triangle insert\\[6pt]
        \includegraphics[scale=0.45]{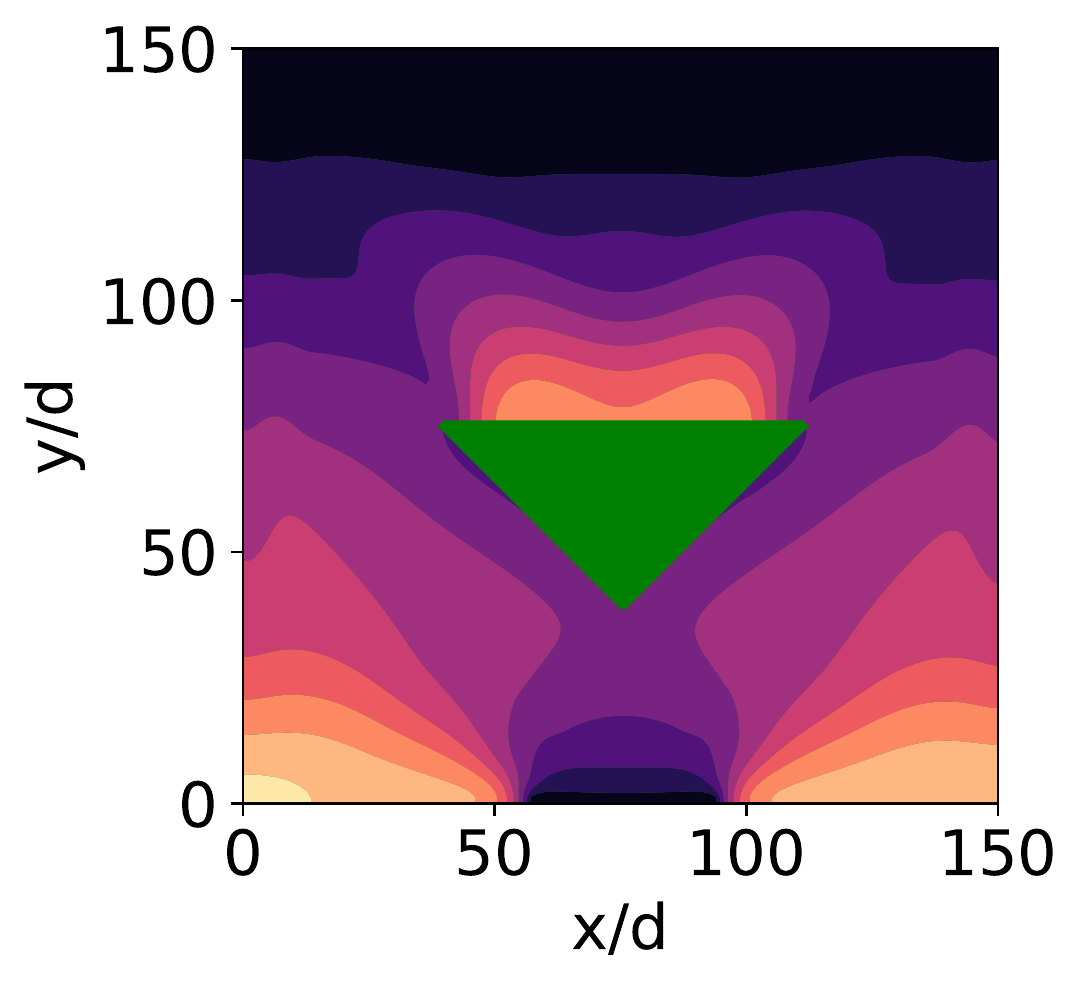} &
        \includegraphics[scale=0.45]{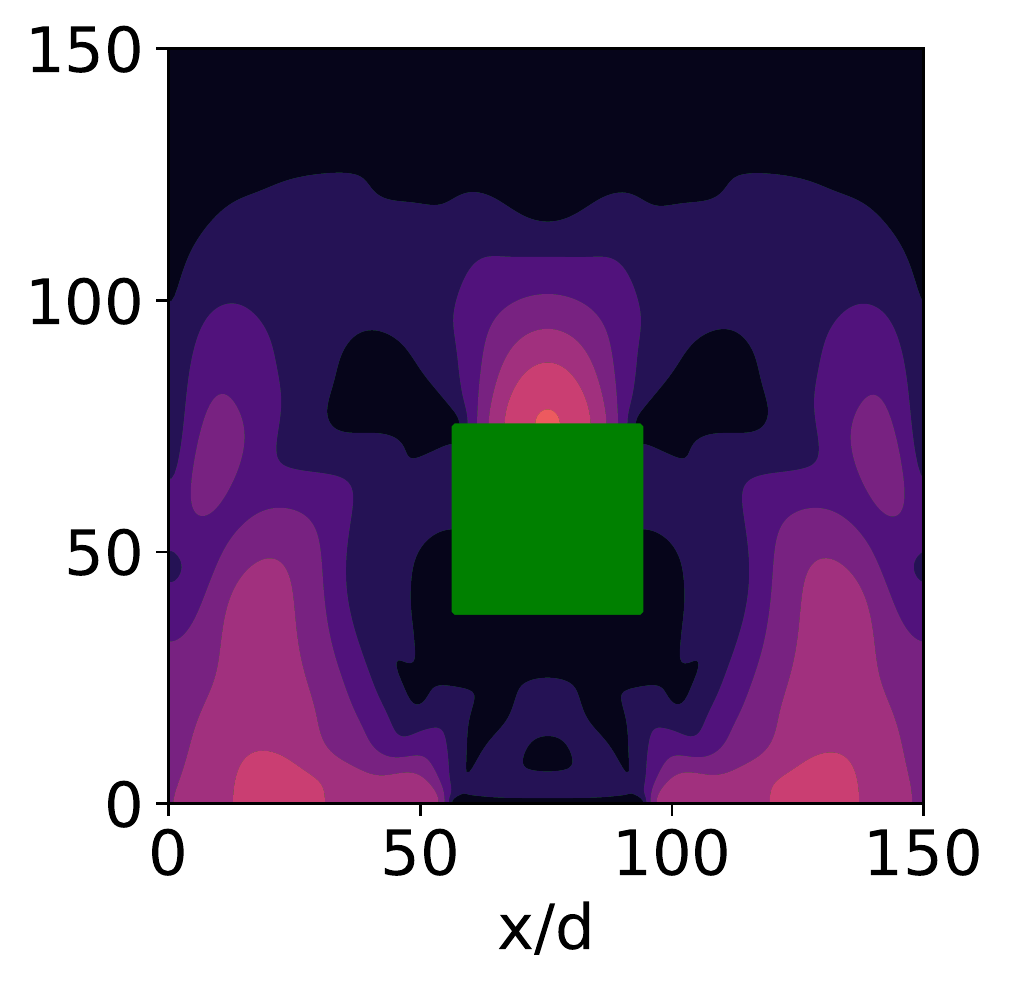} &
        \includegraphics[scale=0.45]{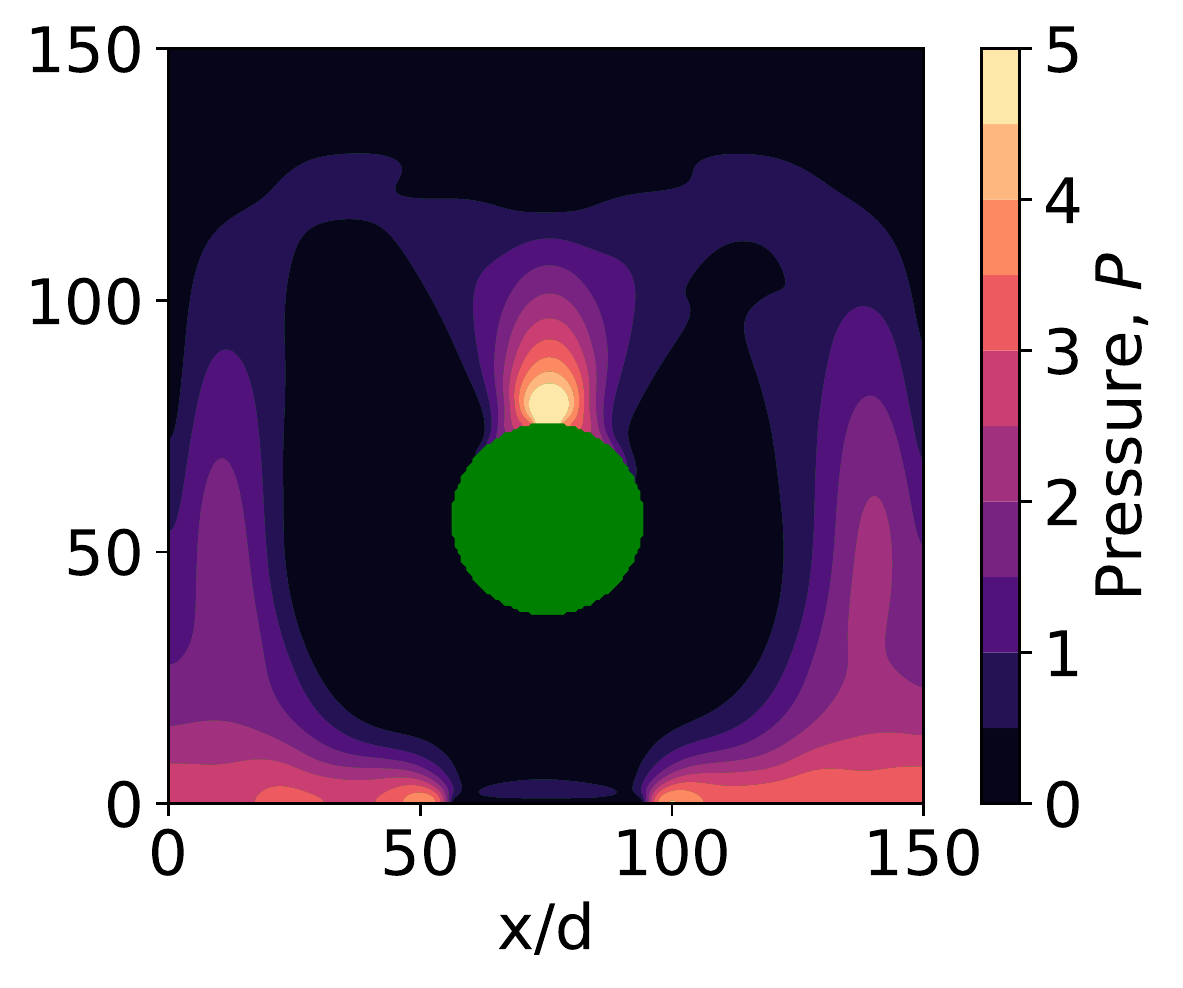} \\
        Downwards triangle insert & Square insert & Circle insert\\[6pt]
    \end{tabular}
    \caption{Granular pressure in silos with various different insert shapes. Pressure is nondimensionalised by $P = P_*/(\rho_* g_*d_*^{-2})$ where $P_*$ is the dimensional granular pressure, $\rho$ is the density of granular material with solid fraction $1$, $g_*$ is gravitational acceleration, and $d$ is the granule diameter. The square, diamond, and circle inserts are sized to have height and width equal to the width of the silo opening, with the triangle inserts having the same height and twice the width.}
    \label{fig:shape_pressure}
\end{figure}

Granular pressure is also important to consider, with contours for each shape shown in Figure~\ref{fig:shape_pressure}. Each of these cases has higher granular pressure in the corners of the silo representing the `hydrodynamic' pressure, with lower pressure in the middle above the orifice. The inserts all seem to widen the low pressure zone. The inserts also have a peak of pressure above the insert itself and a decreased pressure zone directly below the insert.

\begin{figure}
    \centering
    \includegraphics[width=0.7\textwidth]{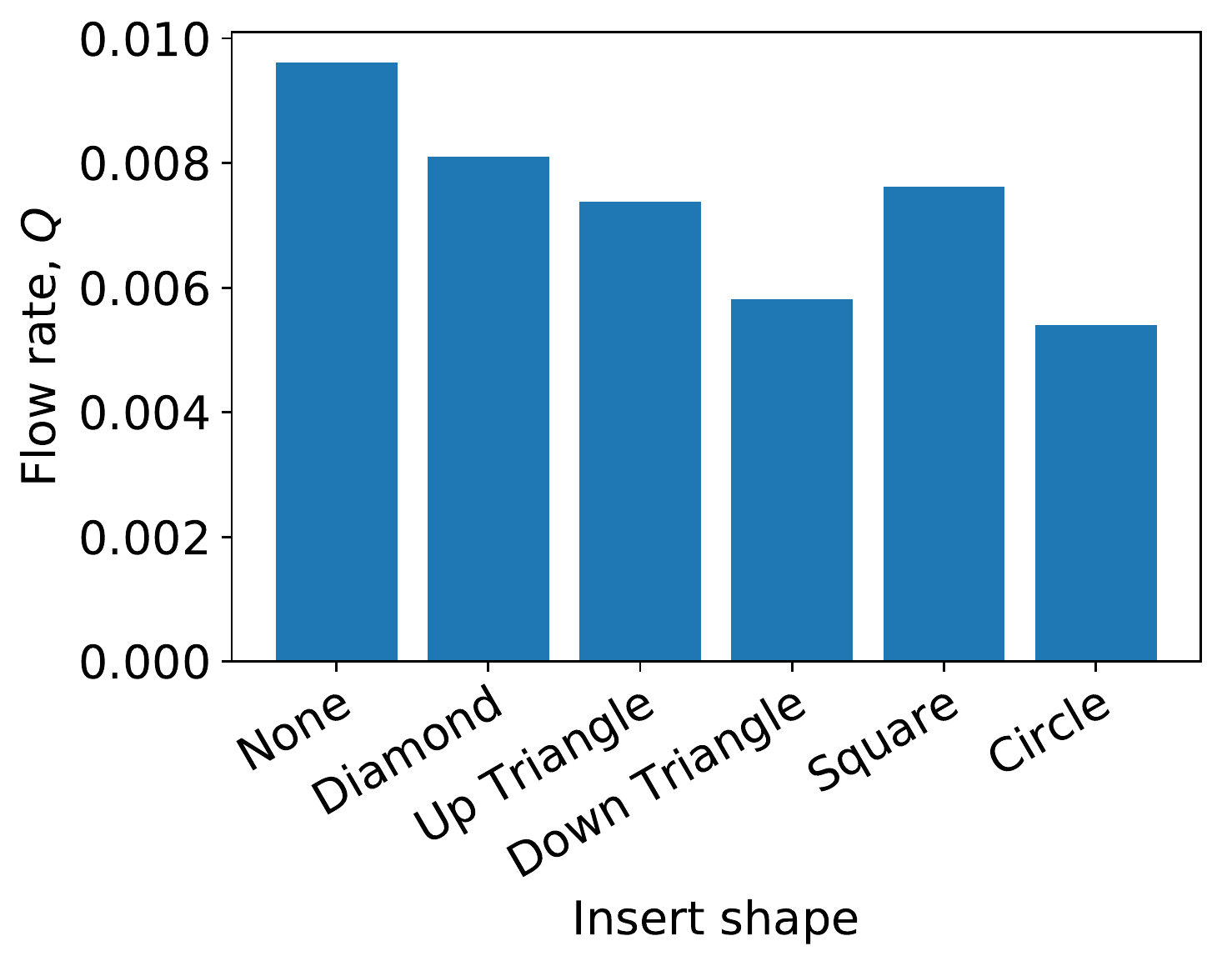}
    \caption{Flow rate $Q=\hat{Q}/\sqrt{gd^5}$ for different shapes of inserts. The square, diamond, and circle inserts are sized to have height and width equal to the width of the silo opening, with the triangle inserts having the same height and twice the width.}
    \label{fig:shape_flow}
\end{figure}

Differently shaped inserts exhibit different flow rates. Finding an insert which reduces the static area while also maintaining a high flow rate may be desirable for industrial applications. The flow rate is calculated by tracking the total mass of material in the silo over time and taking a linear fit. The flow rates for different insert shapes is shown in Figure~\ref{fig:shape_flow}. The case with no insert has the highest flow rate, with all inserts decreasing the flow rate somewhat. Interestingly, the pressure does not seem to have a significant effect on the flow rate, with insert shapes with high pressure in the corners (e.g. the diamond and downward triangle) and low pressure in the corners (e.g. square and circle) giving both relatively high and low flow rates.

\subsubsection{Extensions}
\begin{figure}
    \centering
    \begin{tabular}{ccc}
        \includegraphics[scale=0.45]{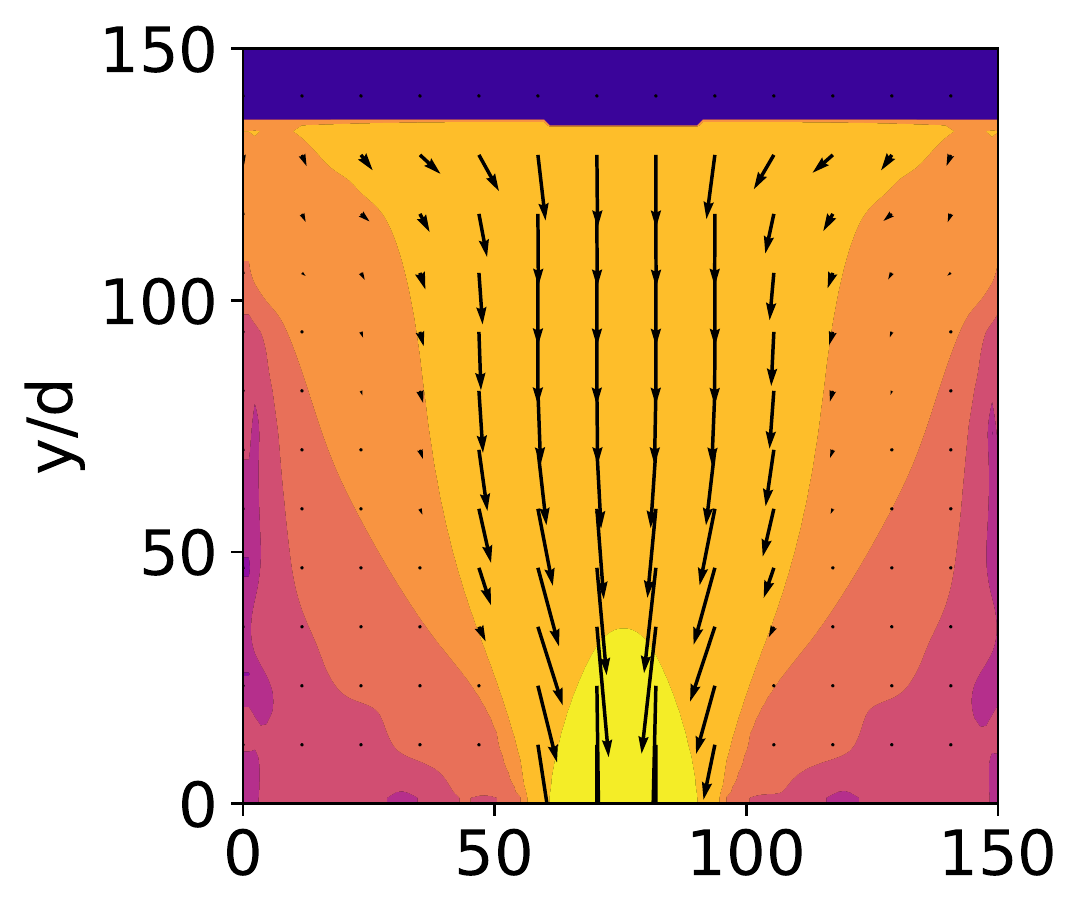} &
        \includegraphics[scale=0.45]{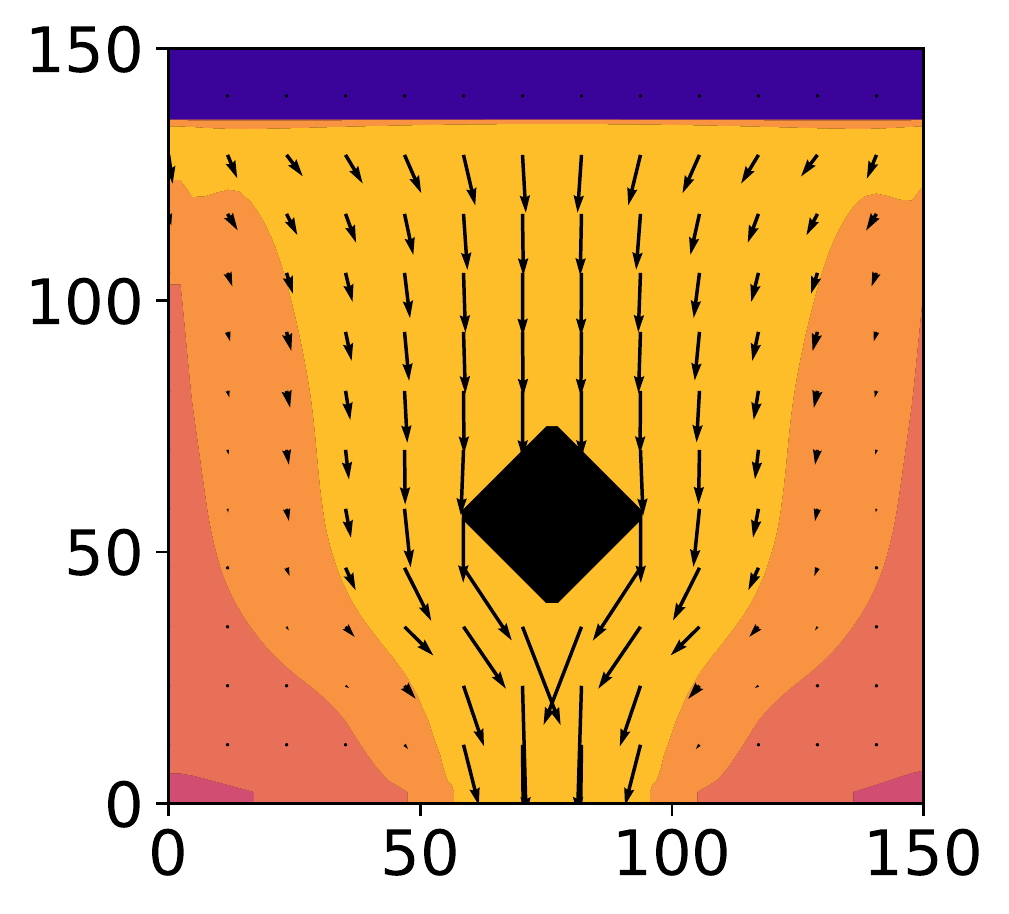} &
        \includegraphics[scale=0.45]{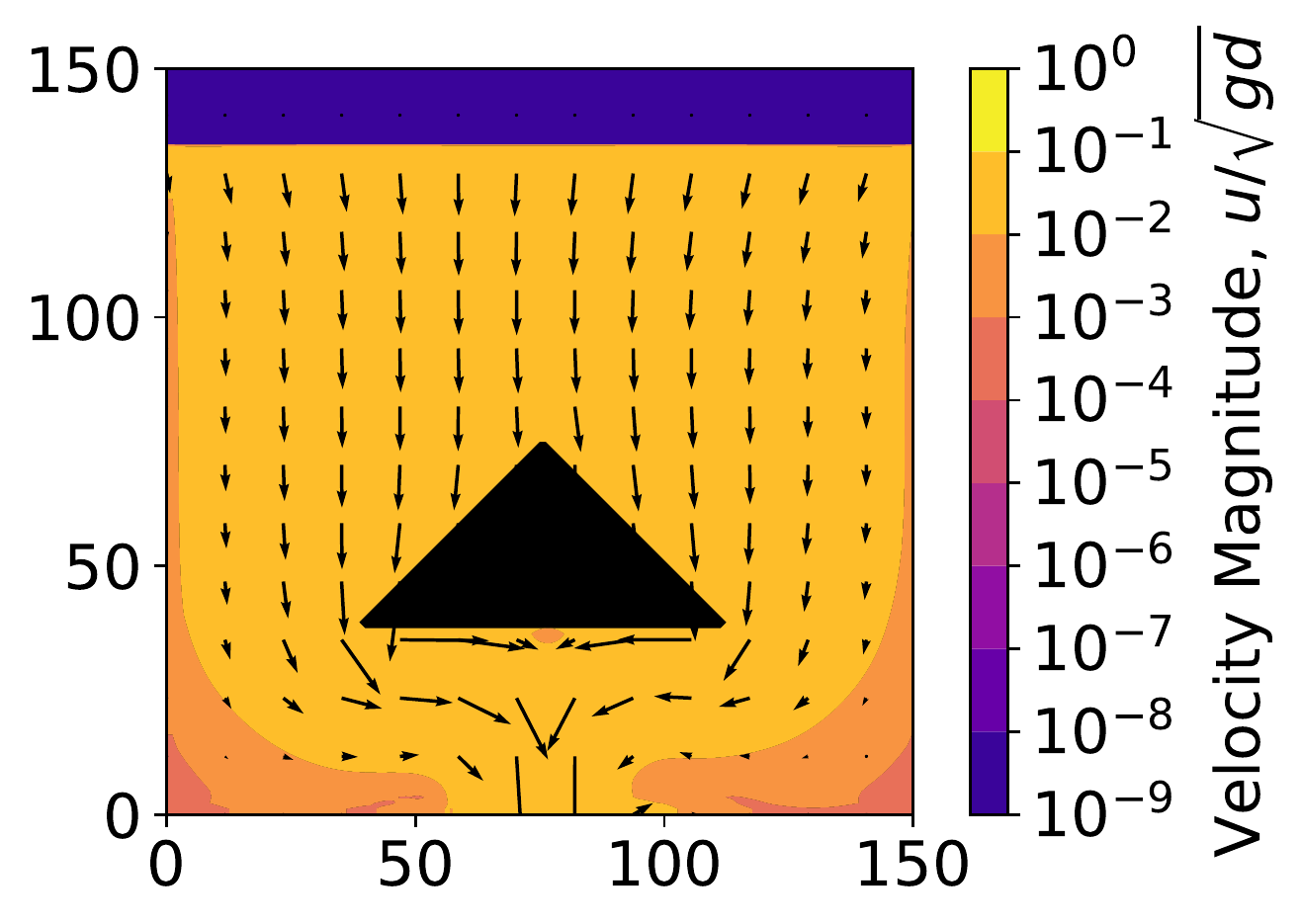} \\
        No insert & Diamond insert & Upwards triangle insert\\[6pt]
        \includegraphics[scale=0.45]{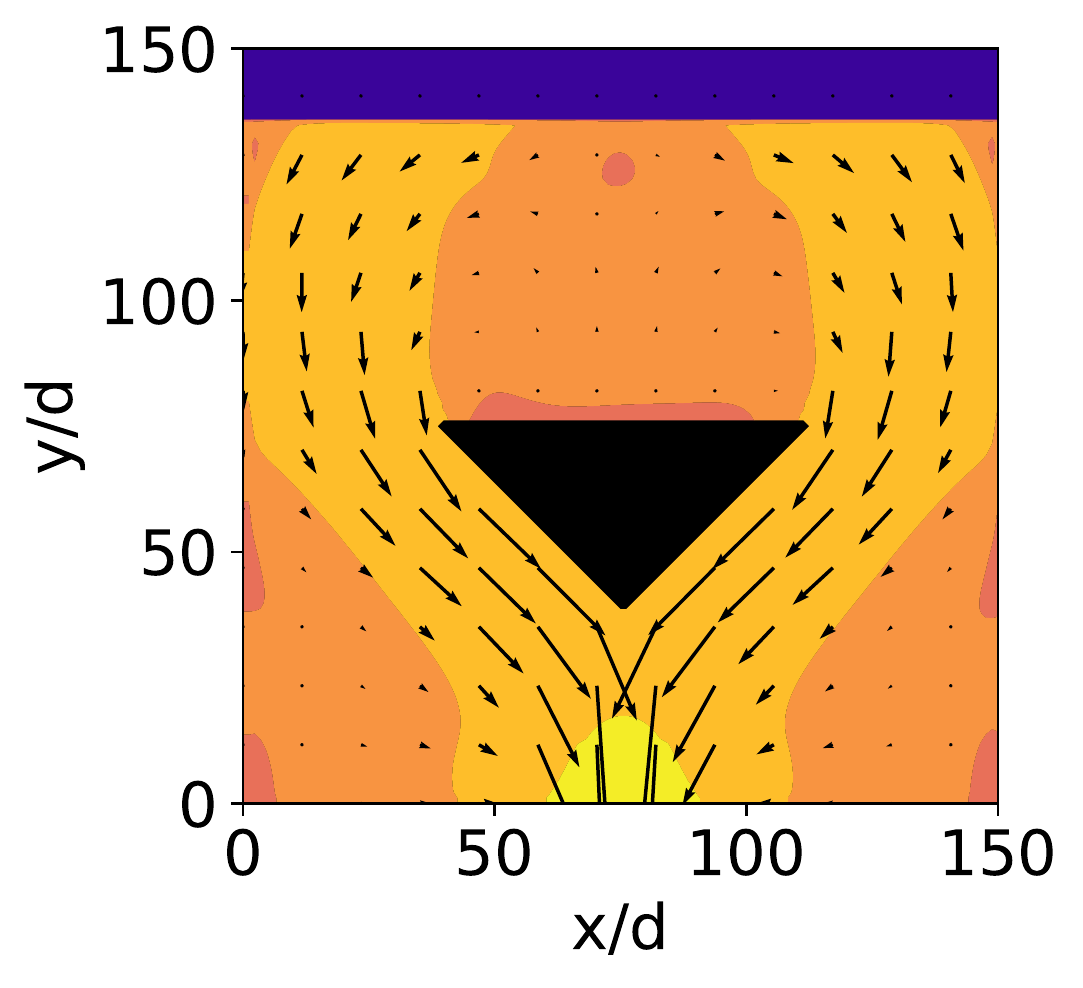} &
        \includegraphics[scale=0.45]{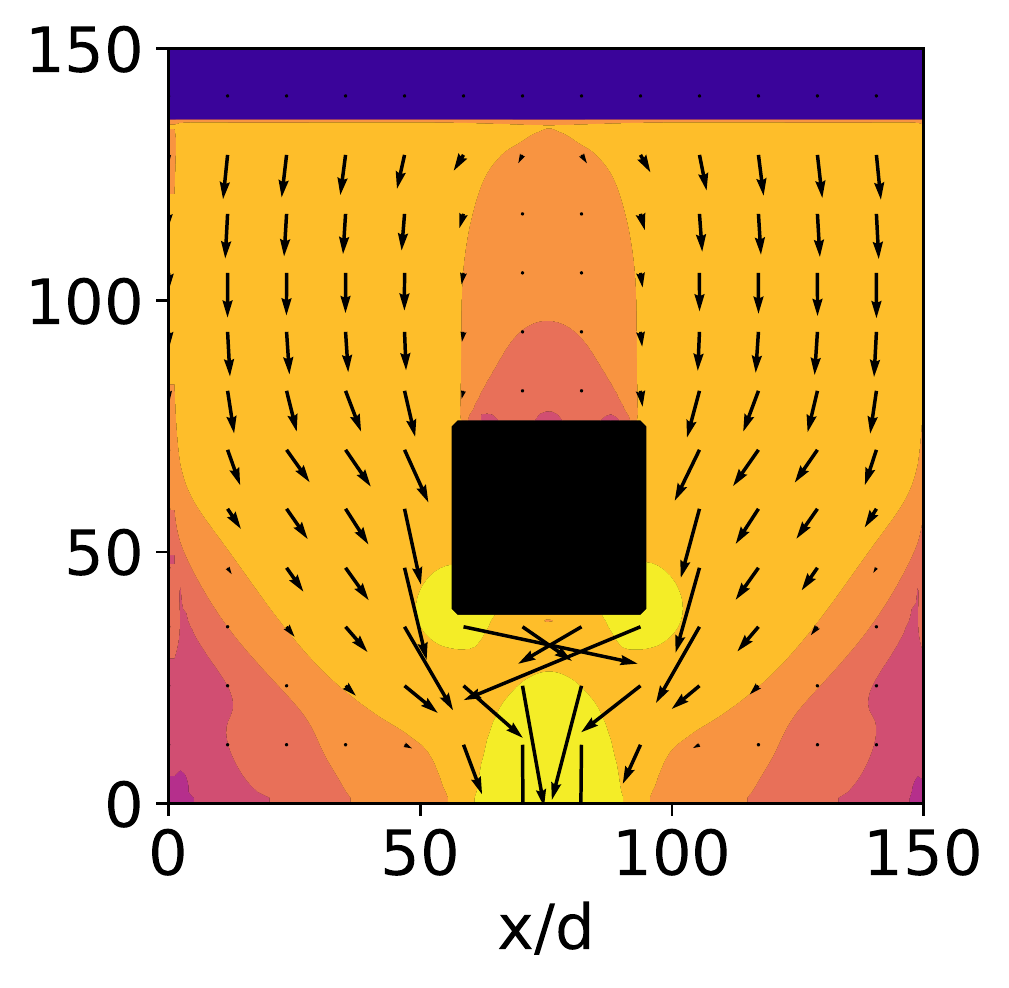} &
        \includegraphics[scale=0.45]{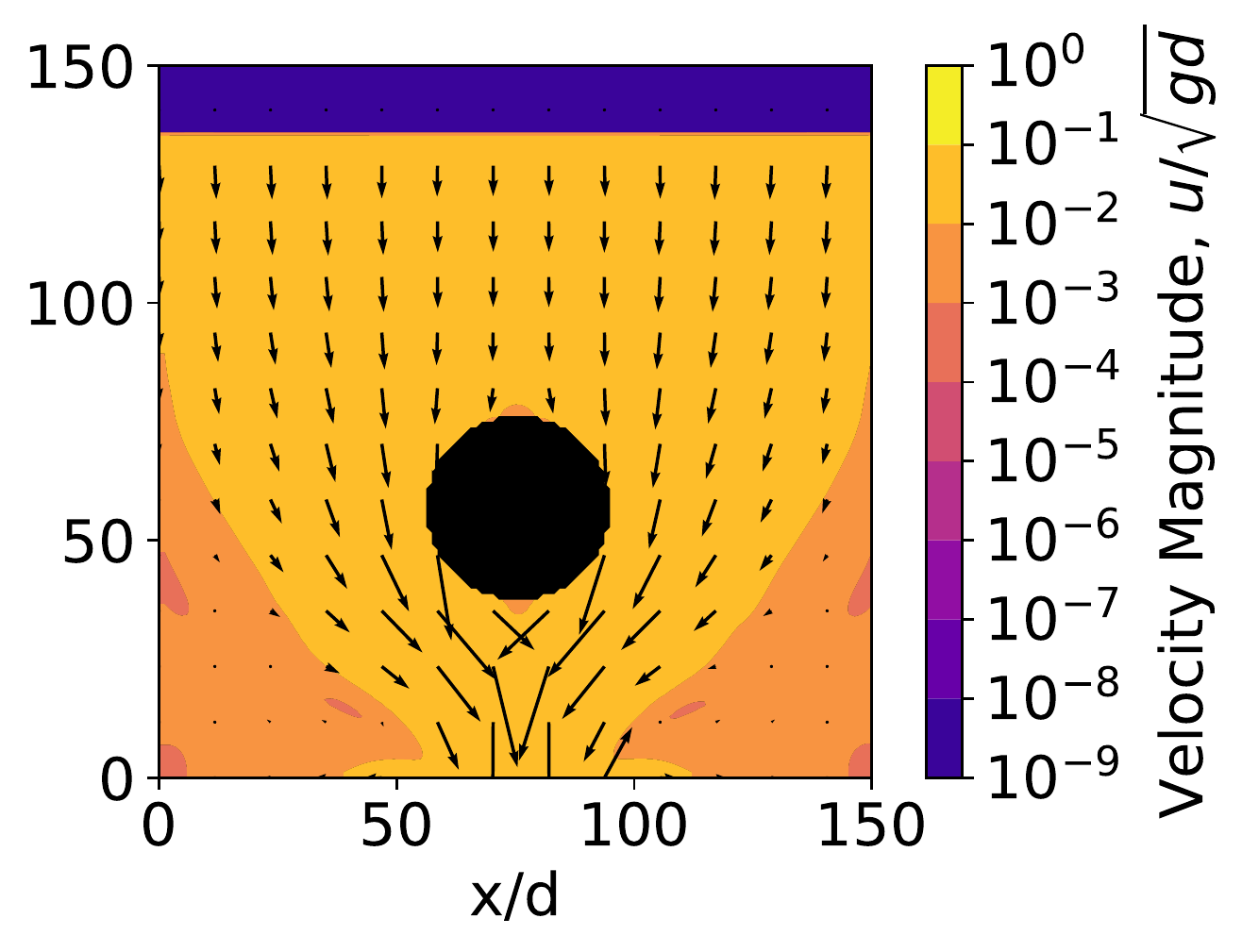} \\
        Downwards triangle insert & Square insert & Circle insert\\[6pt]
    \end{tabular}
    \caption{Velocity in silos with various different insert shapes with dilatancy extension with dilatancy strength $\phi_\textrm{grad}=0.2$. The square, diamond, and circle inserts are sized to have height and width equal to the width of the silo opening, with the triangle inserts having the same height and twice the width.}
    \label{fig:dil_shape_contour}
\end{figure}

\begin{figure}
    \centering
    \begin{tabular}{ccc}
        \includegraphics[scale=0.45]{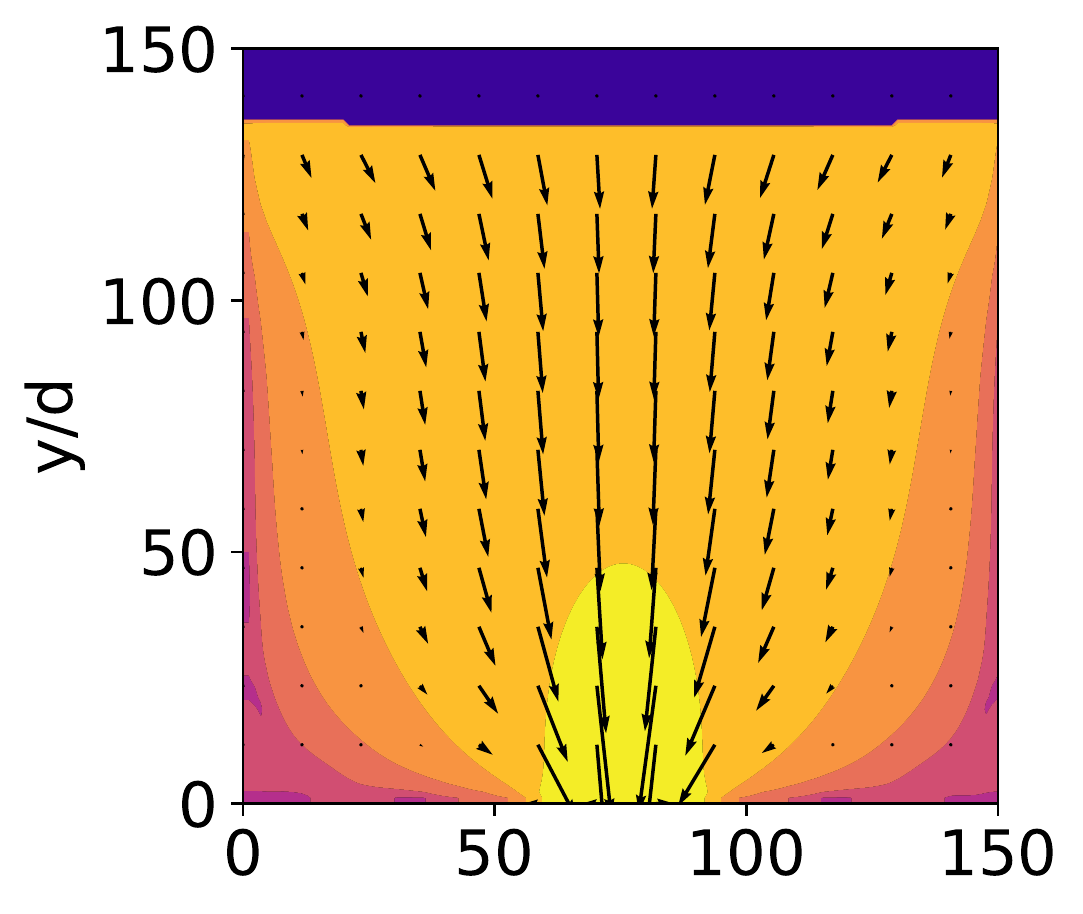} &
        \includegraphics[scale=0.45]{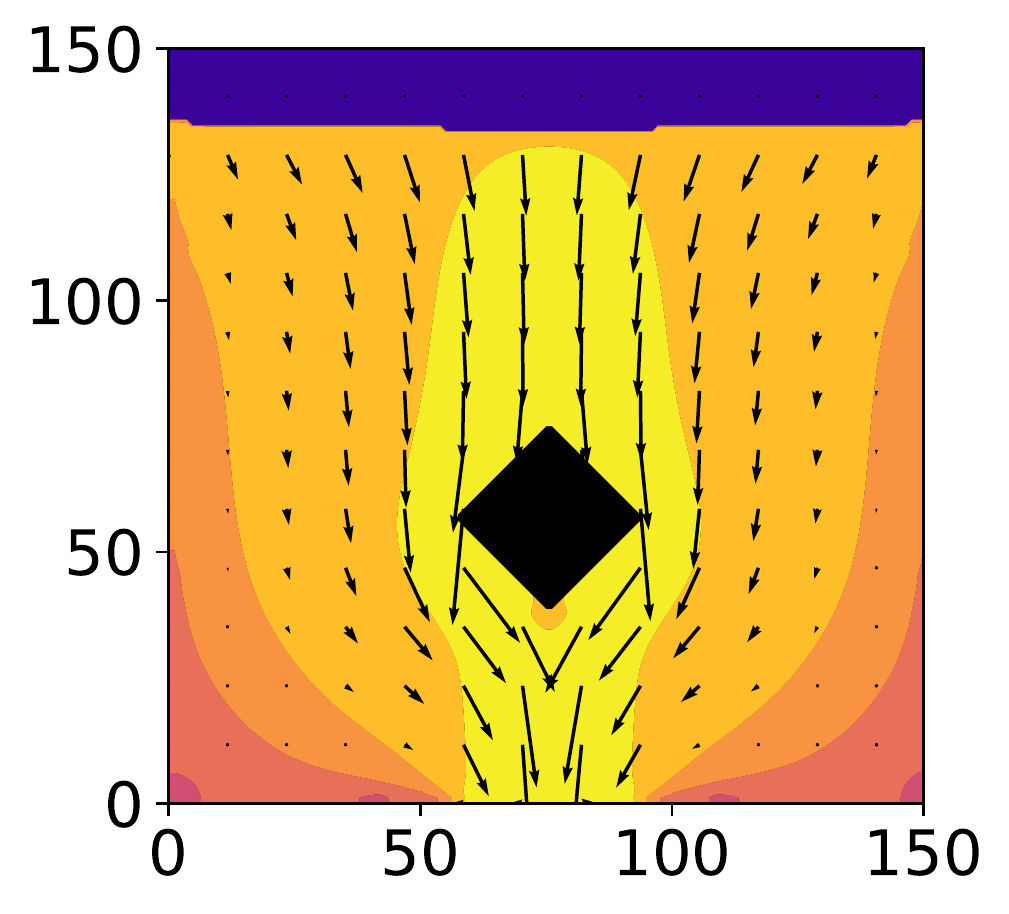} &
        \includegraphics[scale=0.45]{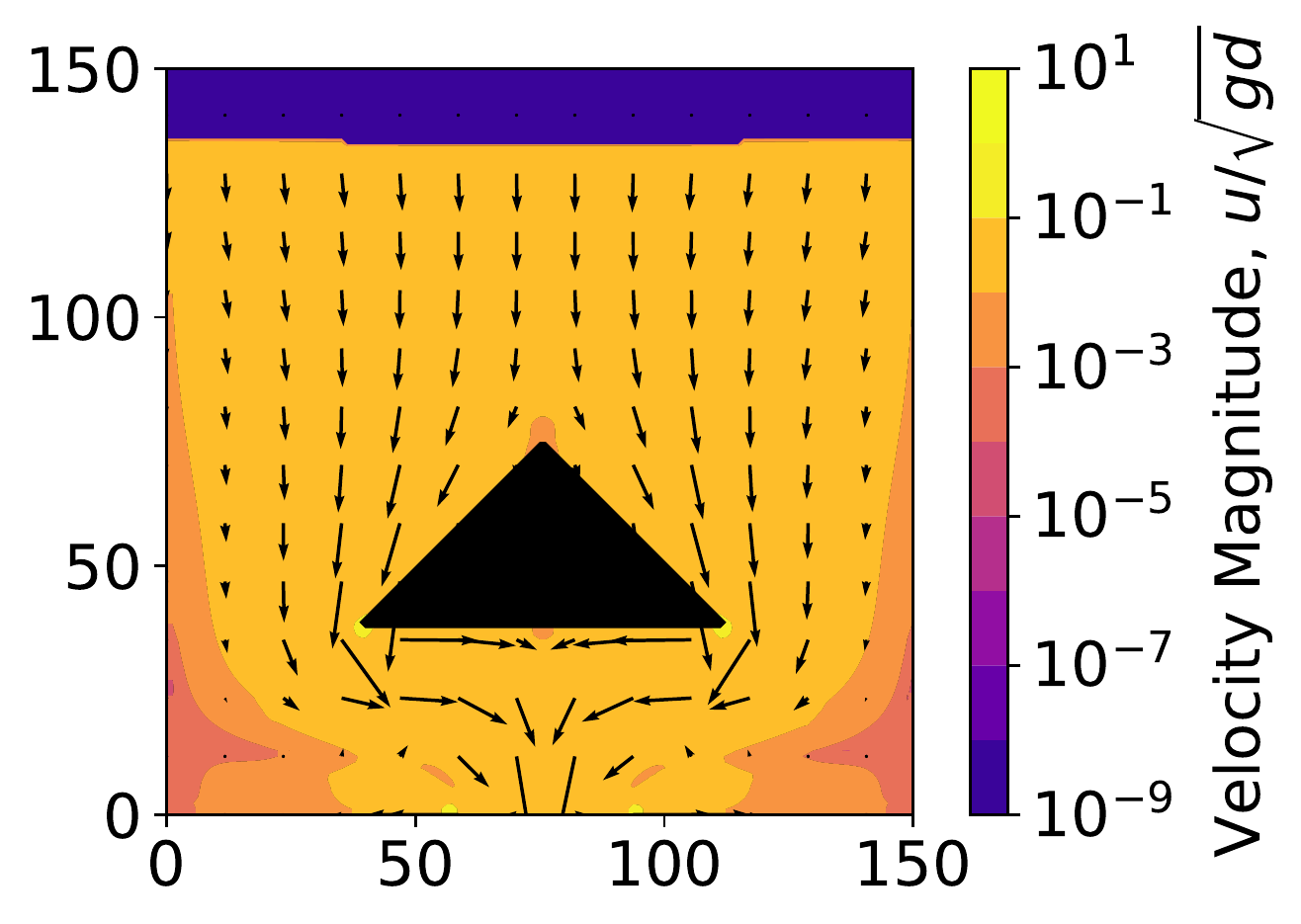}\\
        No insert & Diamond insert & Upwards triangle insert\\[6pt]
        \includegraphics[scale=0.45]{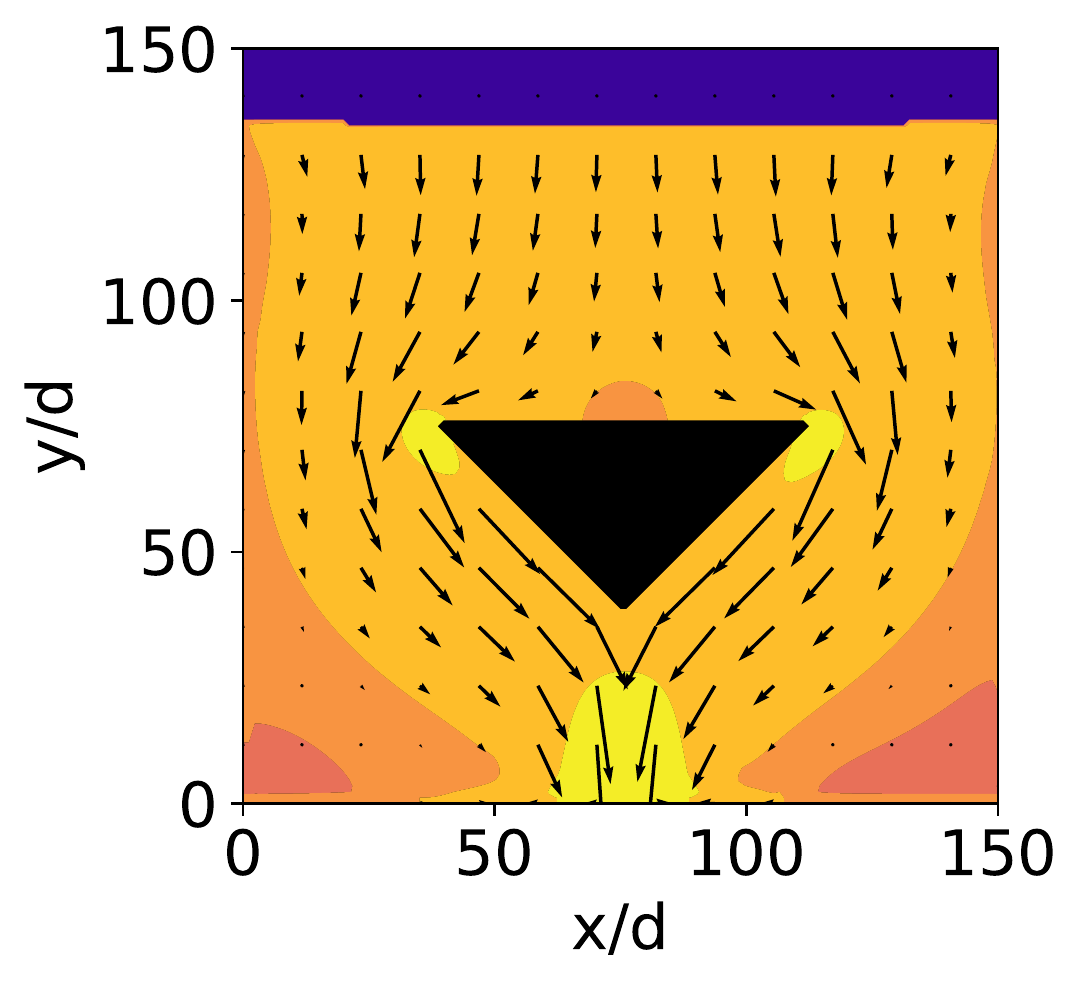} &
        \includegraphics[scale=0.45]{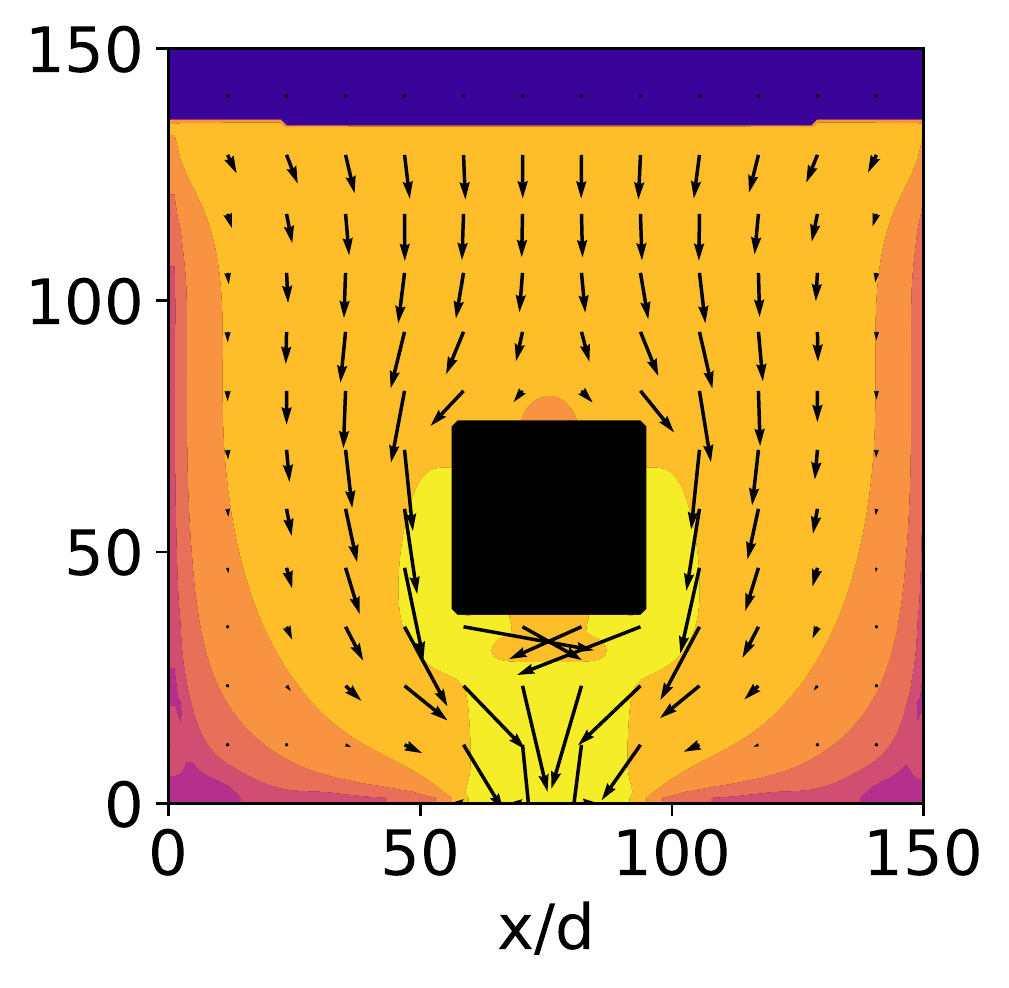} &
        \includegraphics[scale=0.45]{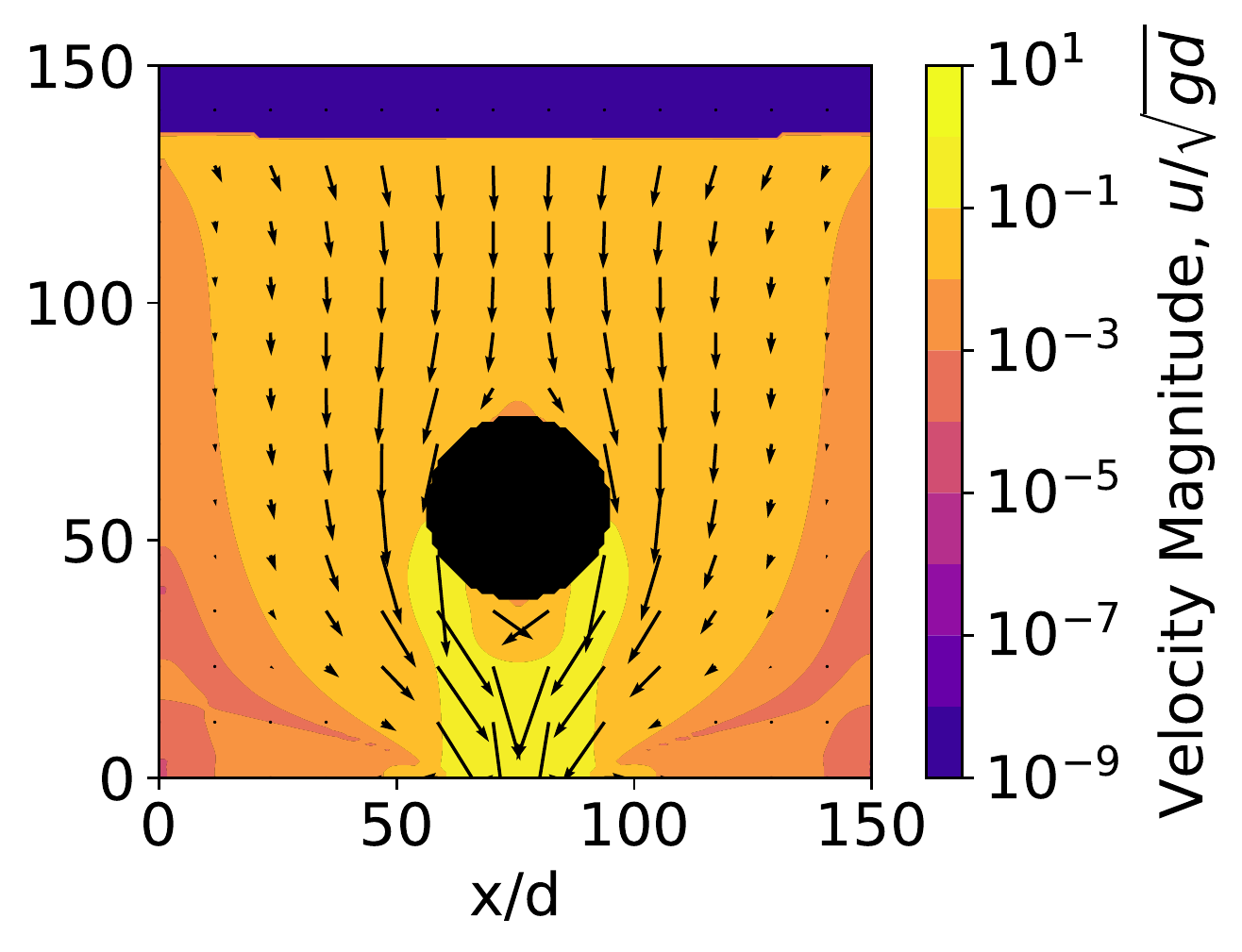} \\
        Downwards triangle insert & Square insert & Circle insert\\[6pt]
    \end{tabular}
    \caption{Velocity in silos with various different insert shapes with nonlocal extension with nonlocal strength $A = 0.5$. The square, diamond, and circle inserts are sized to have height and width equal to the width of the silo opening, with the triangle inserts having the same height and twice the width.}
    \label{fig:nl_shape_contour}
\end{figure}

We can modify the $\mu(I)$ model to capture dilatancy and to capture nonlocal effects either independently or combined. These extensions change the velocity pattern within the silo. Figure~\ref{fig:dil_shape_contour} shows the velocity contour with dilatancy, and Figure~\ref{fig:nl_shape_contour} shows the velocity contour with nonlocal effects. Dilatancy decreases the overall velocity throughout the silo, with peak velocity near the silo outlet less than the peak velocity given in the case without dilatancy. However, dilatancy does not greatly change the direction or shape of the flow pattern. In contrast, nonlocal effects seem to increase the velocity slightly and also changes the shape of the velocity contour. The nonlocal case has slow moving regions that cover most of the areas near the bottom and side walls, reducing the size of the static zones. This may be from the Laplacian in the fluidity model `spreading out' the zero velocity resulting from the no-slip boundary conditions. These low flow areas are displayed in some experiments~\cite{fullard2019dynamics}, so nonlocal effects may be important for capturing flow behaviour near the wall.

\begin{figure}
     \centering
        \includegraphics[width=0.7\textwidth]{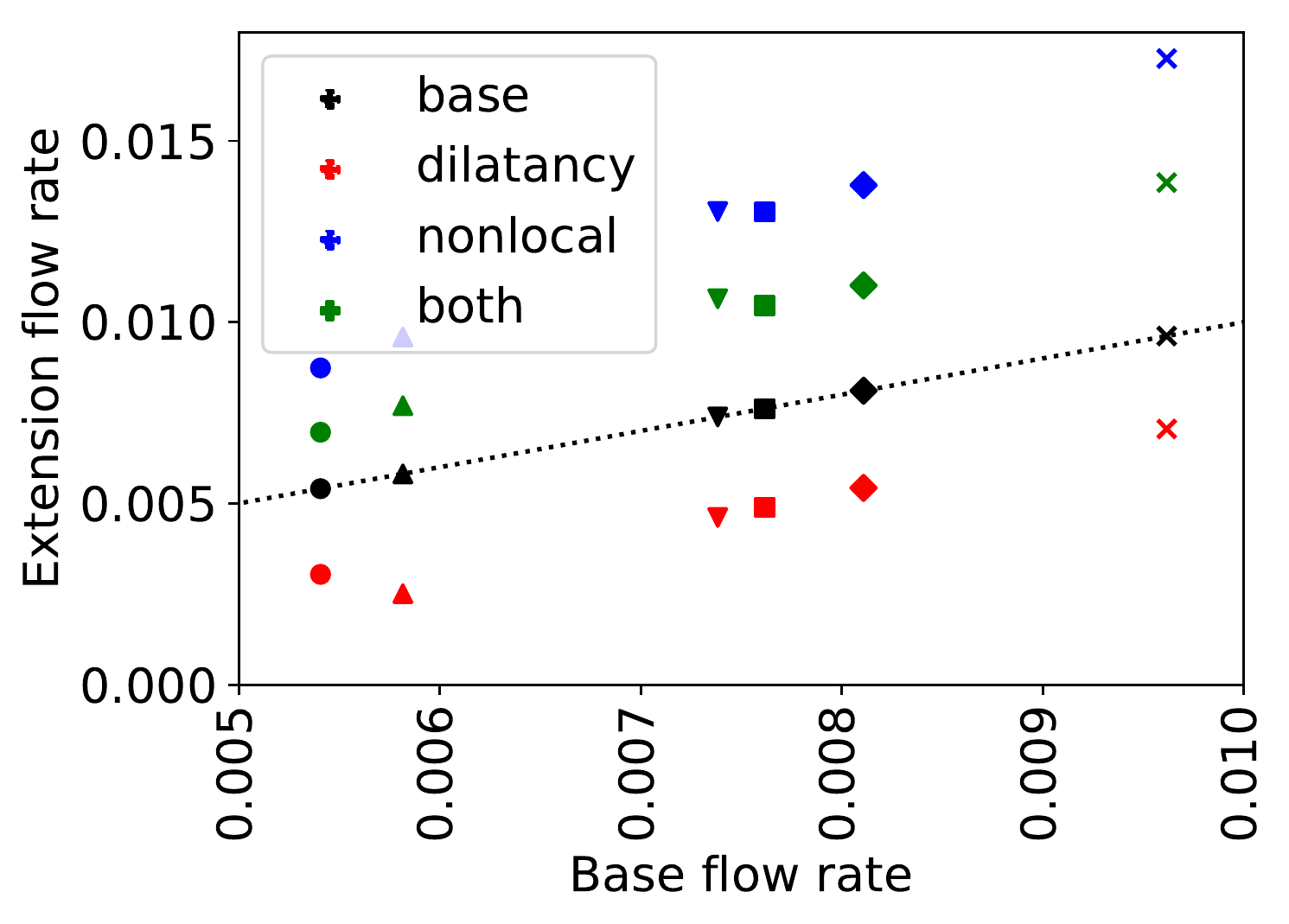}
        \caption{Flow rate $Q=\hat{Q}/\sqrt{gd^5}$ for different extensions, with dilatancy strength $\phi_\textrm{grad}=0.2$ and nonlocal strength $A = 0.5$. The shape of the marker indicates the shape on the insert, with the `x' indicating no insert. The dotted line is the `$x=y$' line, and the base data on this line by definition. The square, diamond, and circle inserts are sized to have height and width equal to the width of the silo opening, with the triangle inserts having the same height and twice the width.}
        \label{fig:extension}
\end{figure}

Figure~\ref{fig:extension} shows the flow rate in the silo, comparing the base $\mu(I)$ model, the model with dilatancy, the model with nonlocal fluidity, and the model with both dilatancy and nonlocal fluidity added. These are also compared for each of the insert shapes studied (indicated by the shape of the data points, with the `x' shape indicating no insert). Nonlocal effects seem to increase the flow rate compared to the base $\mu(I)$ model by a factor of approximately $1.7$, dilatancy decreases the flow rate by a factor of approximately $0.67$, with a combination of both increasing the flow rate by a factor of approximately $1.35$. These factors seems to be fairly consistent across the different shapes, indicating that these effects do not interact with insert shape in a manner that greatly affects the flow rate.

\subsubsection{Static zones}
For a physical flat bottomed silo we would expect static zones in the sides, however the $\mu(I)$ model is incapable of truly static flow. True static regions would have divergent effective viscosity in the static zone. For numerical stability the effective viscosity is capped and zones which should be static will exhibit ``creeping'' flow. Even if this limitation of maximum effective viscosity was removed, physical silos display jamming behaviour, intermittent flows, and other frictional behaviour that this continuous model would not be able to capture. However, we expect the zones which are static in physical silos will also be very low speed in this continuous model. Even if these zones are not truly static, very low speed areas would also result in problems in industrial contexts as this may result in some stored material taking a long time to pass through the silo and potentially deteriorating the quality of the product. As such, we propose that very low speed zones in the continuum model make a suitable analogy for static zones. For the following results we approximate the area of static zones by assuming that any zone that has velocity $u < 10^{-3}\sqrt{gd}$ is static. This static cutoff corresponds to a velocity $<0.1\%$ of the velocity at the opening.

\begin{figure}
     \centering
         \includegraphics[width=0.7\textwidth]{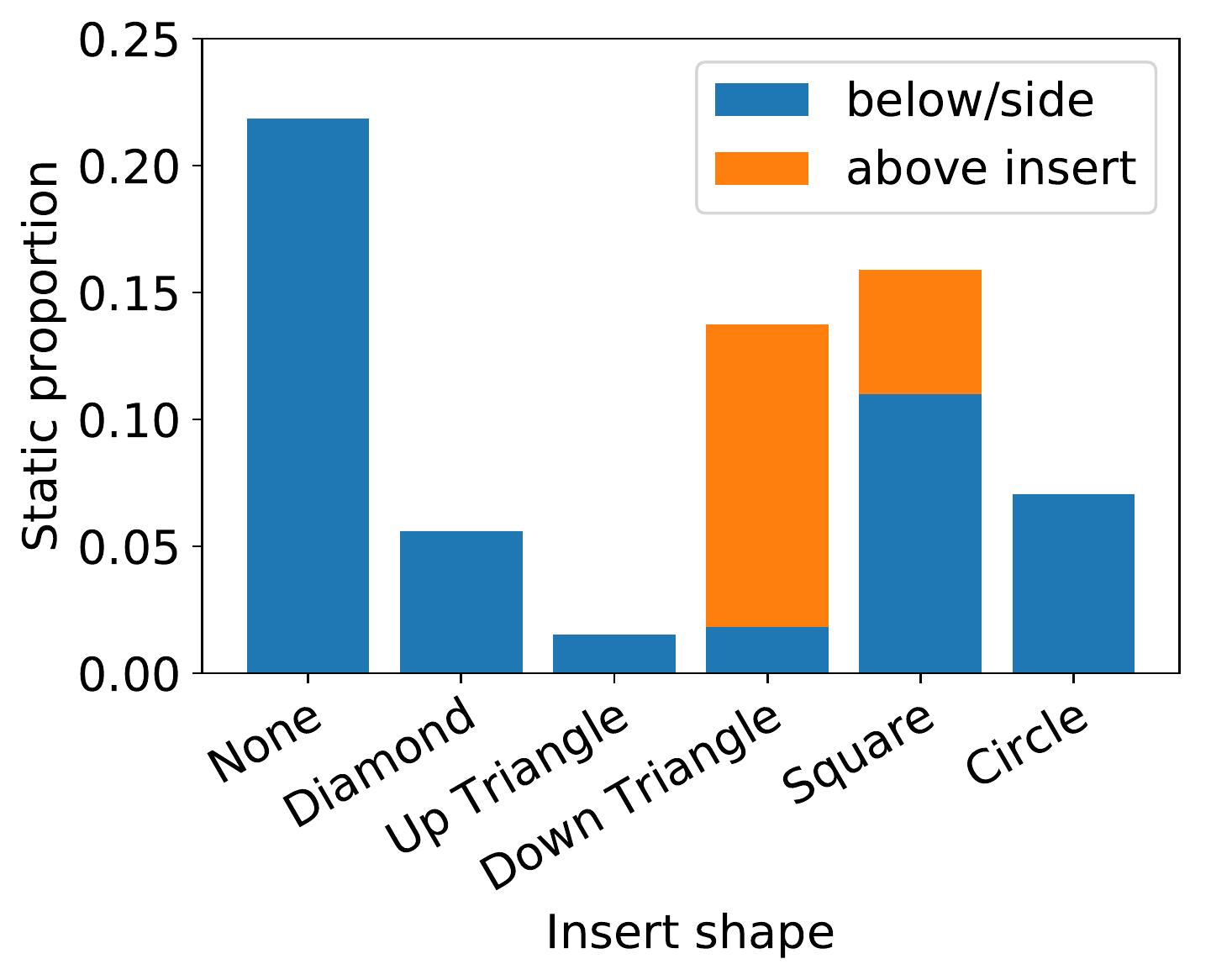}

        \caption{Proportion of the material in the silo that is static by insert shape. The static region is defined as any location where velocity is less than $10^{-3}\sqrt{gd}$. The static material directly above the insert ($y>75d$ and $|x-75d|<75d/2$) is shown in orange, while static material at the sides or base of the silo is shown in blue.}
        \label{fig:shapes_static}
\end{figure}

As shown in Figure~\ref{fig:shape_contour}, inserts change the shape and reduce the size of these static zones. The proportion of the silo containing static material for each shape of silo using the base $\mu(I)$ model is given in Figure~\ref{fig:shapes_static}. As expected, the insert-free case has the largest proportion of the silo filled with static material. Both the downwards pointing triangle and the square have significantly higher static proportions. In part this is due to the flat surface at the top of the insert which creates an additional area where the material can be static. The portion of static material in this region (which we define to be $y>75d$ and $|x-75d|<75d/2$) is shown in orange, while static material outside this region is shown in blue. The flat surface almost completely explains the extra static material for the down triangle, but the square insert has increased static proportion even without this area. The upwards pointing triangle has a particularly low static proportion, this is likely due to the wide base of the triangle forcing flow through the corners of the silo which would otherwise be static.

\begin{figure}
     \centering
        \includegraphics[width=0.7\textwidth]{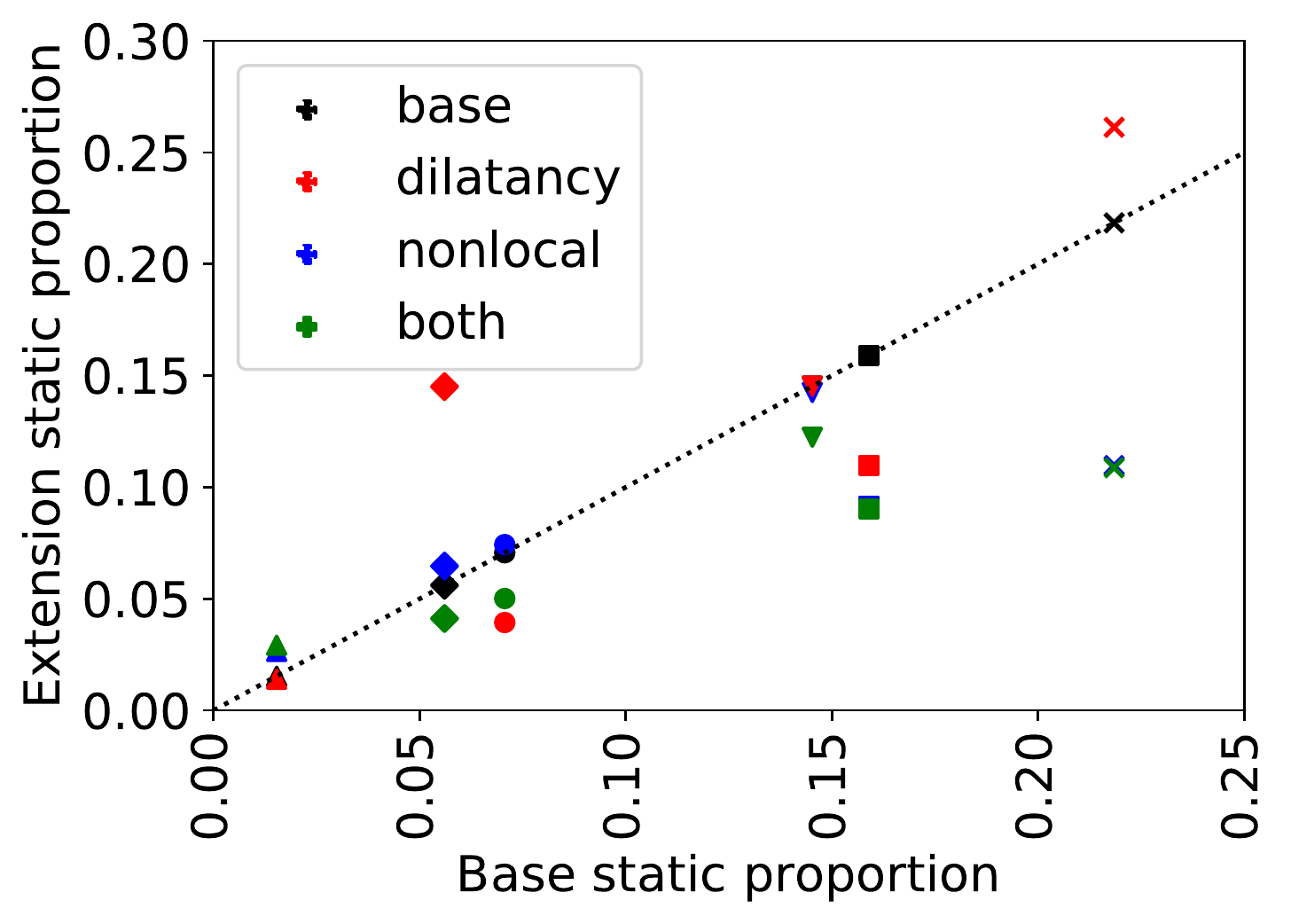}
        \caption{Proportion of the material in the silo that is static for different extensions by insert shape. The static region is defined as any location where velocity is less than $10^{-3}\sqrt{gd}$. The shape of the insert is indicated by the marker shape of the data, with the colour indicating which extensions are enabled. The black points are by definition on the line $x=y$ (the dotted line), with vertical deviation indicating a difference caused by the extension(s).}
        \label{fig:ext_static}
\end{figure}

The impact of extensions on static areas is also important to consider. Figure~\ref{fig:ext_static} shows the comparison of the base $\mu(I)$ model with simulations using combinations of dilatancy and nonlocal extensions. These show a much less consistent pattern than the flow rate shown in Figure~\ref{fig:extension}, with different shapes interacting with extensions differently. Dilatancy seems to increase the area of static material for both diamond and no inserts, but decreases it for circle in square inserts. The nonlocal extension seems to have almost no effect except for the no insert and the square insert case, where it significantly decreases the static area. The presence of both extensions decreases the static area for most tested insert shapes, but particularly for the square insert and no insert cases.

\subsection{Variable insert size}

We have examined inserts of fixed size at a fixed height, however the size and location of inserts will also affect the flow rate and static zones. For examining these parameters we limit ourselves to the diamond shape insert, we vary the height of the insert $H_i$ and the width of the insert $W_i$.

\begin{figure}
    \centering
    \begin{tabular}{cc}
        \includegraphics[width=0.4\textwidth]{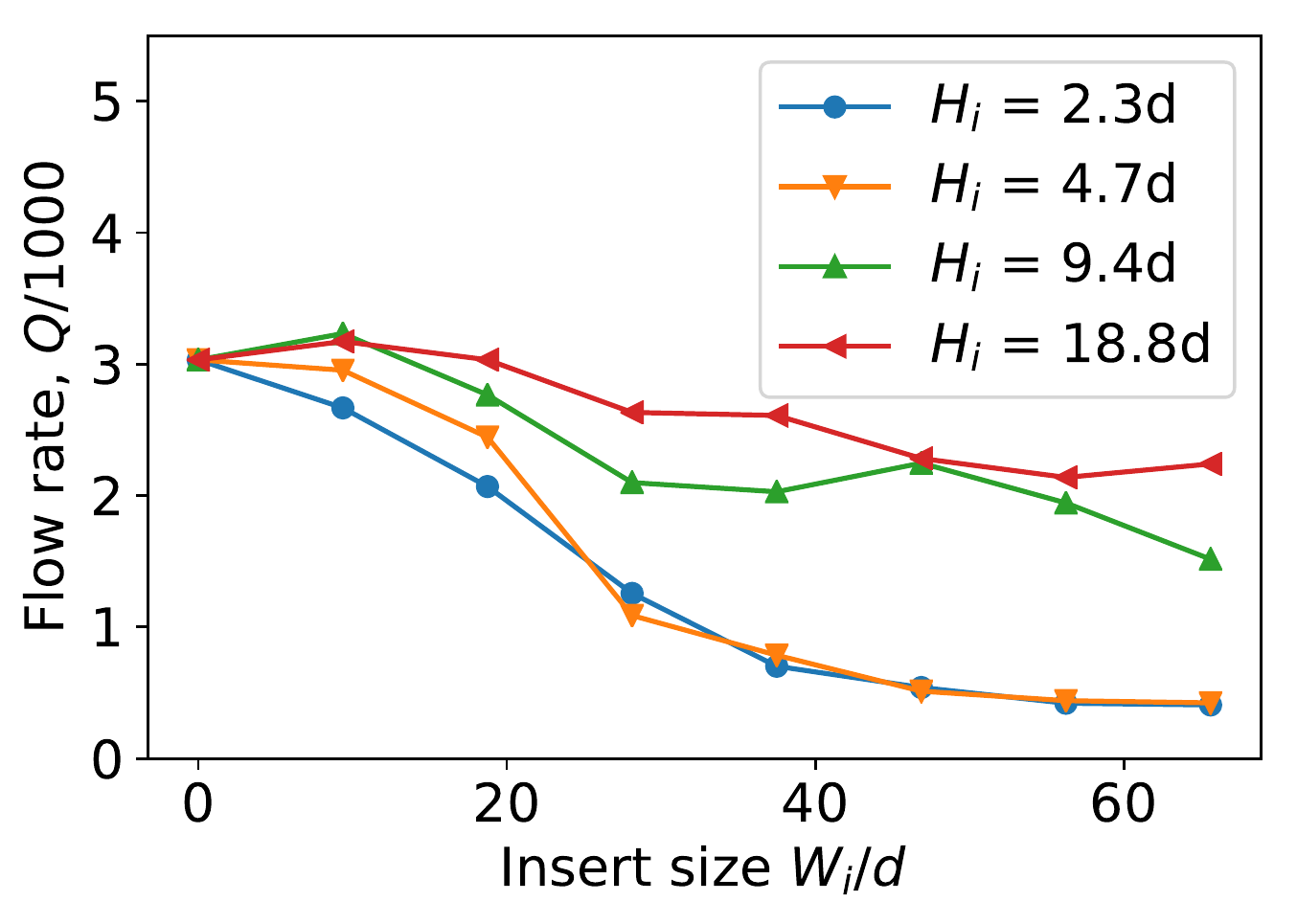} &   \includegraphics[width=0.4\textwidth]{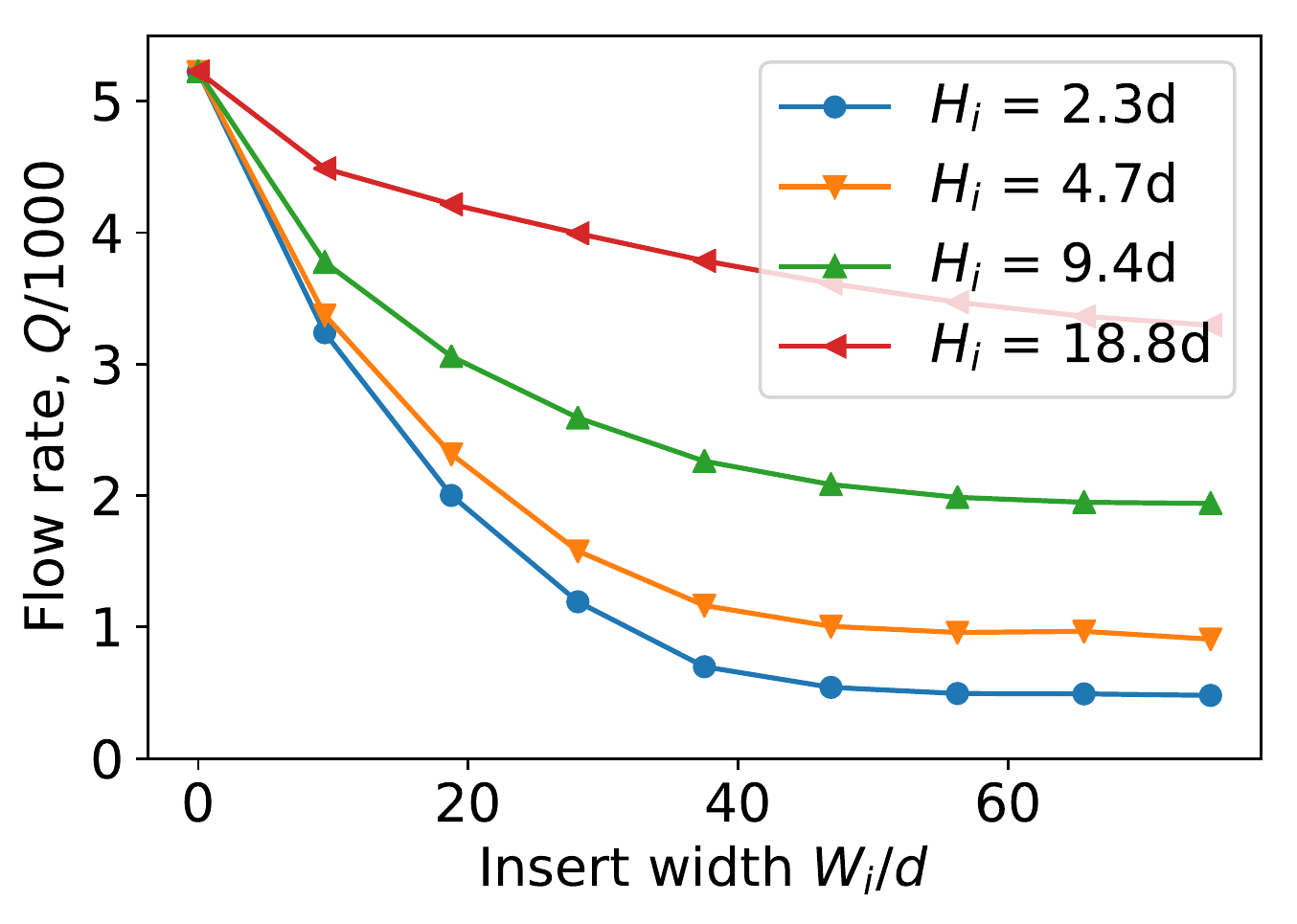} \\
        (a) Local $\mu(I)$ & (b) Nonlocal fluidity\\[6pt]
    \end{tabular}
    \caption{Flow rate $Q=\hat{Q}/\sqrt{gd^5}$ vs different diamond insert sizes at different heights. The base model is shown in (a), with (b) showing the flow rate when nonlocal effects are added.}
    \label{fig:flow_rate}
\end{figure}

We examine the flow rate for variable insert size and location, which is shown in Figure~\ref{fig:flow_rate}. When the insert is small the flow rate is relatively high with a larger insert decreasing the flow rate. This matches our intuition of a large obstacle near the orifice reducing flow more than a small obstacle far away from the orifice. This effect is stronger when the insert is located closer to the orifice. This matches the intuitive idea that the dynamics of the region near the orifice are the most important for determining flow rate.

For small inserts far away from the orifice in the local $\mu(I)$ model, the flow rate is actually increased compared to the silo with no insert. This may be because when the insert is far away it forces the flow to happen from the sides of the silo, which effectively mobilises more material hence increases the amount of material attempting to approach the orifice. This increased flow is not seen when nonlocal effects are included. The flow rate of nonlocal silos is already higher than that of local silos, so the benefit of more material being `mobilised' by the insert might already be provided by the nonlocal model spreading out the flow, explaining the difference between the local and nonlocal cases.

\begin{figure}
    \centering
    \begin{tabular}{cc}
        \includegraphics[width=0.4\textwidth]{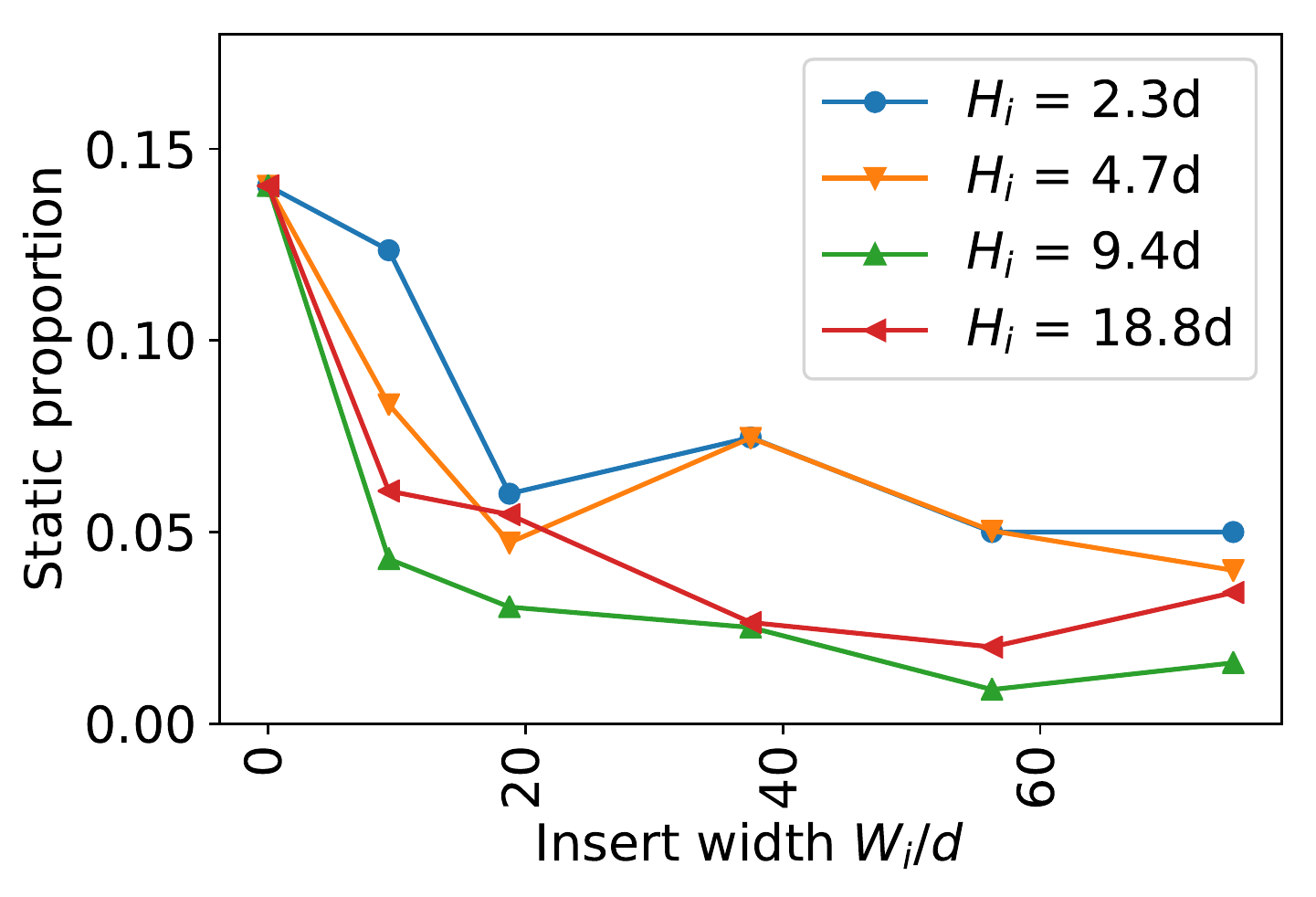} &   \includegraphics[width=0.4\textwidth]{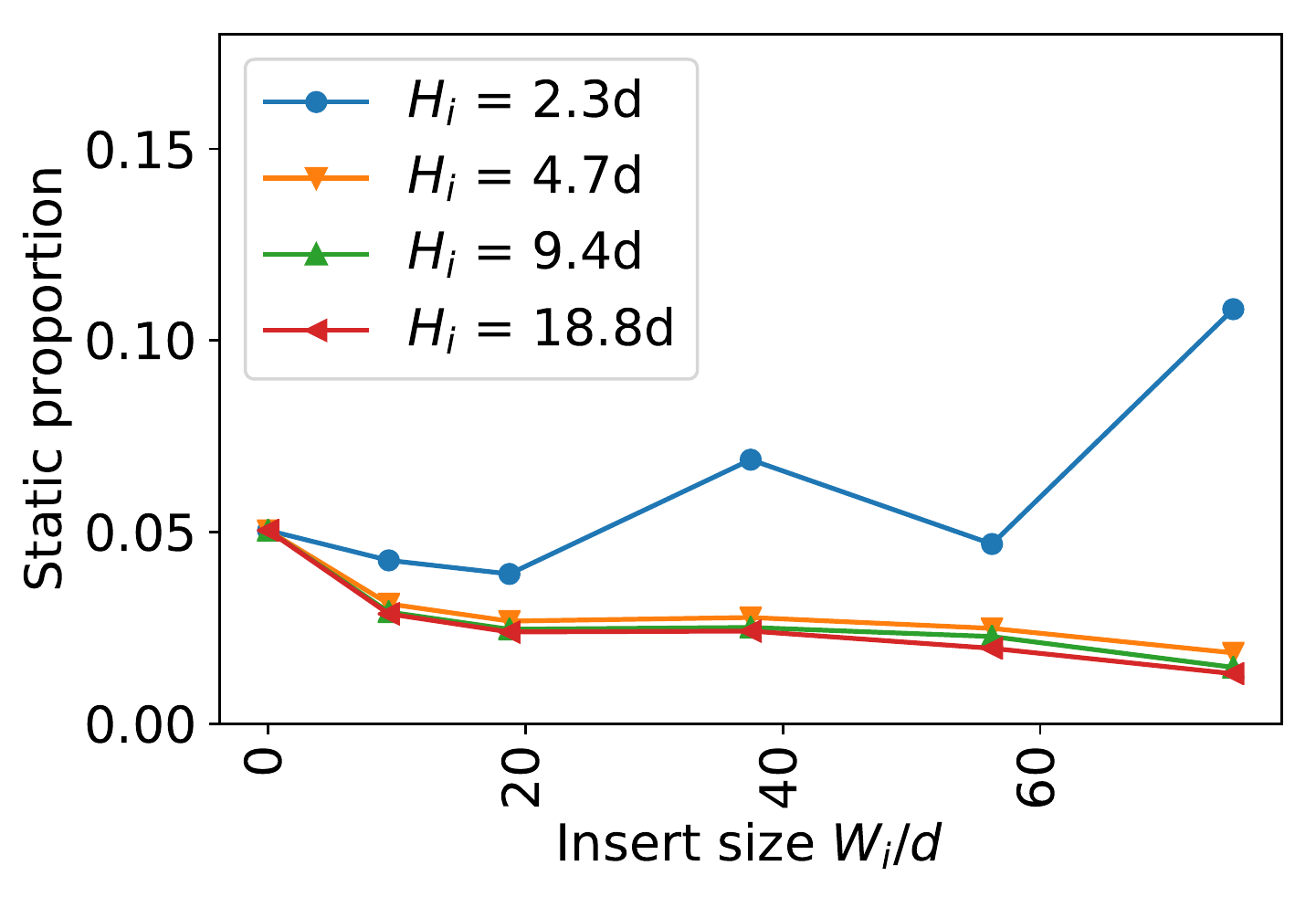} \\
        (a) Local $\mu(I)$ & (b) Nonlocal fluidity\\[6pt]
    \end{tabular}
    \caption{Static region vs different diamond insert sizes at different heights. The base model is shown in (a), with (b) showing the flow rate when nonlocal effects are added.}
    \label{fig:static_size}
\end{figure}

The static zones proportion is shown in Figure~\ref{fig:static_size}. Looking at the base case, for small widths (i.e. $W_i\approx10d$), the height has a large impact on the static area, with inserts at the orifice showing almost no reduction in static area over no insert. For larger inserts the height is less important but still impactful, with lower inserts giving larger static areas up to $H_i\approx20d$, with inserts higher than $H_i\approx20d$ decreasing the impact. This demonstrates that the insert has a maximum effective height for reducing static area. Above this height, increasing the width of the insert only increases flowing area by a small amount. The nonlocal case for large insert heights shows a similar decreasing trend with insert size, but with less volatility than the base case. The overall static values are also a lot lower since the fluidity works to spread out which areas are flowing. However for inserts close to the outlet, increased insert size actually increases static area. The flow rate in this case also becomes very low, indicating that the flow could be moving too slow to make a meaningful distinction between static and non-static material using this model. For the nonlocal model the flow rate depends on the insert height, while the static zone is relatively independent of insert height. This implies that we can achieve static zone reduction with minimal flow rate reduction by using inserts high in the silo.


\section{Conclusion}
We have tested the $\mu(I)$ model with dilatancy and nonlocal effects in silos with inserts. This model and extensions are capable of capturing information about the flow rate and static zones of granular material around various different shapes and sizes of inserts in a silo.

This model shows that extensions have a significant effect on flow rate, with nonlocal effects increasing the flow and dilatancy decreasing flow. Different shapes of inserts also change the flow rate behaviour significantly. Each of these factors seem to be independent on each other, with different combinations of extensions having approximately the same effect on flow rate for a circular insert as for a diamond insert or any other shape of insert. By contrast, the size of static area can change dramatically with different combinations of extensions and insert shapes. This shows that nonlocal effects and dilatancy are vital to calculating static zones, and any model wishing to accurately capture the effect on static zones would be highly dependent on these extensions.

We also examined insert size, showing that there is a reasonably sized insert with both greatly reduces static areas without dramatically reducing flow rate, with larger inserts reducing flow rate without improving static behaviour significantly. The examination of the diamond insert indicates that when the inserts is $~10d$ above the opening and the insert is $~20d$ wide for this silo, the optimal combination of static zone reduction without flow rate reduction is achieved. This information could be used to optimise silo flow in cases where static or slow moving material is a concern.

\bibliographystyle{unsrt} 
\bibliography{aa_ref}

\end{document}